\documentclass[usegraphicx,usenatbib]{mn2e}
\usepackage{bm}
\usepackage{amsmath}
\usepackage{epsfig}
\usepackage{booktabs}
\usepackage{multirow}

\begin{document}

\title[Faraday Rotation and CMB maps]{Faraday Rotation as a diagnostic of Galactic foreground contamination of CMB maps}

\author[M. Hansen et al.]{M. Hansen$^1$, W. Zhao$^1$, A. M. Frejsel$^1$, P. D. Naselsky$^1$, \newauthor J. Kim$^1$ and O. V. Verkhodanov$^2$\\
$^1$Niels Bohr Institute and DISCOVERY center, Blegdamsvej 17, 2100 Copenhagen {\O}, Denmark \\
$^2$Special Astrophysical Observatory, Nizhnij Arkhyz, Karachaj-Cherkesia, 369167, Russia}

\maketitle

\begin{abstract}
The contribution from the residuals of the foreground can have a significant impact on the temperature maps of the Cosmic Microwave Background (CMB). Mostly, the focus has been on the galactic plane, when foreground cleaning has taken place. However, in this paper, we will investigate the possible foreground contamination, from sources outside the galactic plane in the CMB maps. \\
We will analyze the correlation between the Faraday rotation map and the CMB temperature map. The Faraday rotation map is dependent on the galactic magnetic field, as well as the thermal electron density, and both may contribute to the CMB temperature. We find that the standard deviation for the mean cross correlation deviate from that of simulations at the $99.9\%$ level. Additionally, a comparison between the CMB temperature extrema and the extremum points of the Faraday rotation is also performed, showing a general overlap between the two. Also we find that the CMB Cold Spot is located at an area of strong negative cross correlation, meaning that it may be explained by a galactic origin. \\
Further, we investigate nearby supernova remnants in the galaxy, traced by the galactic radio loops. These super nova remnants are located at high and low galactic latitude, and thus well outside the galactic plane. We find some correlation between the Faraday Rotation and the CMB temperature, at select radio loops. This indicate, that the galactic foregrounds may affect the CMB, at high galactic latitudes
\end{abstract}


\section{Introduction}
The study of the cross-correlations between the Faraday rotation and the Cosmic Microwave Background (CMB) temperature, was started in \citep{dineen2004}, and has showed itself to be of remarkable importance in the field of CMB science.\\
The first, third, fifth and seventh year data releases from the Wilkinson Microwave Anisotropy Probe (WMAP) \citep{wmapresults}, \citep{wmap3ytem}, \citep{wmap5ytem}, \citep{wmap7ytem} provide the most precise measurements of the CMB temperature distribution. The precision of these measurements will be significantly improved with the coming data from the ongoing PLANCK mission. In order to be used for precision cosmology, the CMB data must be cleaned from the residuals of the galactic foreground. The focus for this cleaning has been the galactic plane, as this includes the most powerful contributions from the galaxy. However, in this article, we will investigate the possible foreground contamination at high and low galactic latitude. In particular, we will look at the effect that the interstellar medium (ISM), as well as nearby supernova remnants, might have on the CMB.\\
The turbulent ISM consist of hot ($T_g>10^5 K$) low density ($n_e\sim 10^{-3} cm^{-3}$) ionized gas, filling about $70-80\%$ of the Galactic volume and interspersed by cold neutral and relatively dense clouds with a high density of HI atoms ($\la 10^{-2}cm^{-3}$).These HI regions are the places for intensive star formation, which can turn the corresponding HI clouds into the short lived HII zones. The corresponding transition was observed in the blue compact galaxy Henize 2-10, by detection of a compact ($<8 pc$) ~1 mJy radio sources \citep{kobulnicky}. If similar HI-HII transitions occurs in the Milky Way, these clouds could manifest themselves in the CMB range of frequencies as a feature of the optical depth for the CMB photons, creating negative spots in the directions of the clouds, due to the Sunyaev–Zeldovich effect.\\
While the influence of the ISM clouds on the CMB signal could have profound implication for the CMB-cosmology and the ISM-science, this aspect of the problem attract very little attention in the literature. There is global mass balance by evaporation, ionization and condensation between supernova energy injection and radiative cooling in the ISM (see for instance, \citep{mckee}, \citep{melioli}, \citep{stil} and corresponding references therein).\\
Additionally, the galactic sky contains several radio spurs, which can be joined together into loops. These are called the galactic radio loops, and since the 70s, six loops have been identified, commonly referred to, as loop I-VI (for a review see \citep{Borka}). Loop I-IV are designated as the main radio loops, and were discovered first \citep{Large1}, \citep{Quigley}, \citep{Large2}, whereas loops V and VI, were only confirmed later \citep{Milogradov}. The origin for the radio loops are the quasi-spherical remnants of very old, nearby supernovae explosions, and the ensuing shock fronts (see for instance \citep{Urosevic}). The shock front contains an envelope of HI atoms, and violent mixing of warm and hot gasses takes place inside the shock area. The subsequent synchrotron emission from these supernova remnants (SNR), at the galactic radio loops may contribute to the CMB-radiation in form of the foreground residuals. We will investigate this, by analyzing correlations between the magnetic field strength and the CMB temperature, in the loop areas.\\
The goal of our paper is to show that the HI-HII clouds can interfere with the CMB signal, and to investigate a possible connection between the galactic radio loops and the CMB temperature. To test these ideas we looked at the cross-correlations between a WMAP temperature patch of the sky $T(\vec{r})$ and the Faraday rotation depth $\Phi (\vec{r})$, using recently available data \citep{Faraday_depth}.\\
The outline of the paper is the following. First we introduce the concept of Faraday depth, as well as our cross correlation coeffecient, for circular areas on the sky. Then we introduce the two datasets that we use, the data for the Faraday depth, and for the CMB, as well as our masking procedure. In the next section we construct the map of correlations, and present areas of extreme correlations, as well as investigate connections between Faraday depth and CMB temperature extremes. We hereafter compare the correlation maps to simulations, analyzing the significance of our findings. We then turn to the specific area around the Cold Spot, and investigate the significance of the correlations in the area, using various tests.
Next, we start on the analysis of the loop areas in the sky, and look at the correlations there between CMB temperature and Faraday depth, as well as between polarized data and Faraday depth. Finally, in the last section, we summmarize the paper, and draw our conclusions.

\section{CMB temperature-Faraday rotation depth cross-correlation.}
The first test which we would like to implement, is the comparison of the CMB signal and the depth of the Faraday rotation $\Phi$ for various areas of the sky. Historically, the cross-correlations of the CMB and the Faraday rotation map was detected in \citep{dineen2004} for selected patches of the sky, mainly due to incompleteness of the last one. In a recent paper \citep{Faraday_depth}, the most extensive catalog of Faraday rotation data of compact extragalactic polarized radio sources from the VLA sky survey NVSS \citep{nvss}, has been assembled.\\
The authors \citep{Faraday_depth} introduce the concept of Faraday depth, which depends on position and is independent of any astrophysical source. The Faraday depth, corresponding to a position at a distance $r_0$ from an observer, is given by a line of sight integral,
\begin{eqnarray}
\Phi (r_0) = \frac{e^3}{2\pi m^2_e c^4} \int_{r_0}^{0} dr n_e (r) B_e(r)=\frac{e^3}{2\pi m^2_e c^4}I(\theta,\phi)
\label{eq4}
\end{eqnarray}
over the thermal electron density $n_e$ and the line of sight component of the magnetic field $B_r$. Here, $e$ and $m_e$ are the electron charge and mass, $c$ is the speed of light, and $\theta$ and $\phi$ are the polar and azimuthal angles of the polar system of coordinates. From Eq.\ref{eq4} one can see a remarkable feature of the function $I(\theta,\phi)$. If the magnetic field is nearly constant within some coherence lenght $\lambda_B$, the Faraday depth $\Phi(\theta,\phi)$ and 
$I(\theta,\phi)$ are proportional to the Thompson optical depth $\tau(\theta,\phi)$. \\
In order to look for correlations between the CMB temperature measurements and Faraday rotation measure, we define the cross-correlation map $r_s(l, b, R)$ as follows
\begin{eqnarray}\label{smap}
r_s(l,b,R) \! = \! \frac{\sum_{i\in C}(x_i \! - \! \bar{x})(y_i \! - \! \bar{y})}{\sqrt{\sum_{i\in C}(x_i \! - \! \bar{x})^2 \! + \! \epsilon}\sqrt{\sum_{i\in C}(y_i \! - \! \bar{y})^2 \! + \! \epsilon}},
\end{eqnarray}
where $\epsilon=10^{-6}$, is used to avoid the possible divergence in the calculation. $C$ is the circular region around $(l,b)$ with radius being $R$. $\bar{x}$ and $\bar{y}$ are the mean values of $x_i$ and $y_i$ in the kq75p1 masked maps. In our analysis, $x_i$ denotes the Faraday map and $y_i$ denotes the CMB temperature map.
It is clear that $-1<r_s<1$, and $r_s > 0$ indicates the positive correlation while $r_s < 0$ shows a negative correlation. 
If $R=360^{\circ}$, $r_s(l,b,R)$ becomes independent of $(l,b)$, and reduces to the well-defined full-sky correlation coefficient. When $R\ll360^{\circ}$, $r_s(l,b,R)$ describes the local correlation of Faraday rotation map and CMB map around the location $(l,b)$ with scale $R$.

\section{Data used in the analysis}

\subsection{Galactic Faraday rotation map}
\begin{figure}
  \begin{center}
    \centerline{\includegraphics[width=1.0\linewidth]{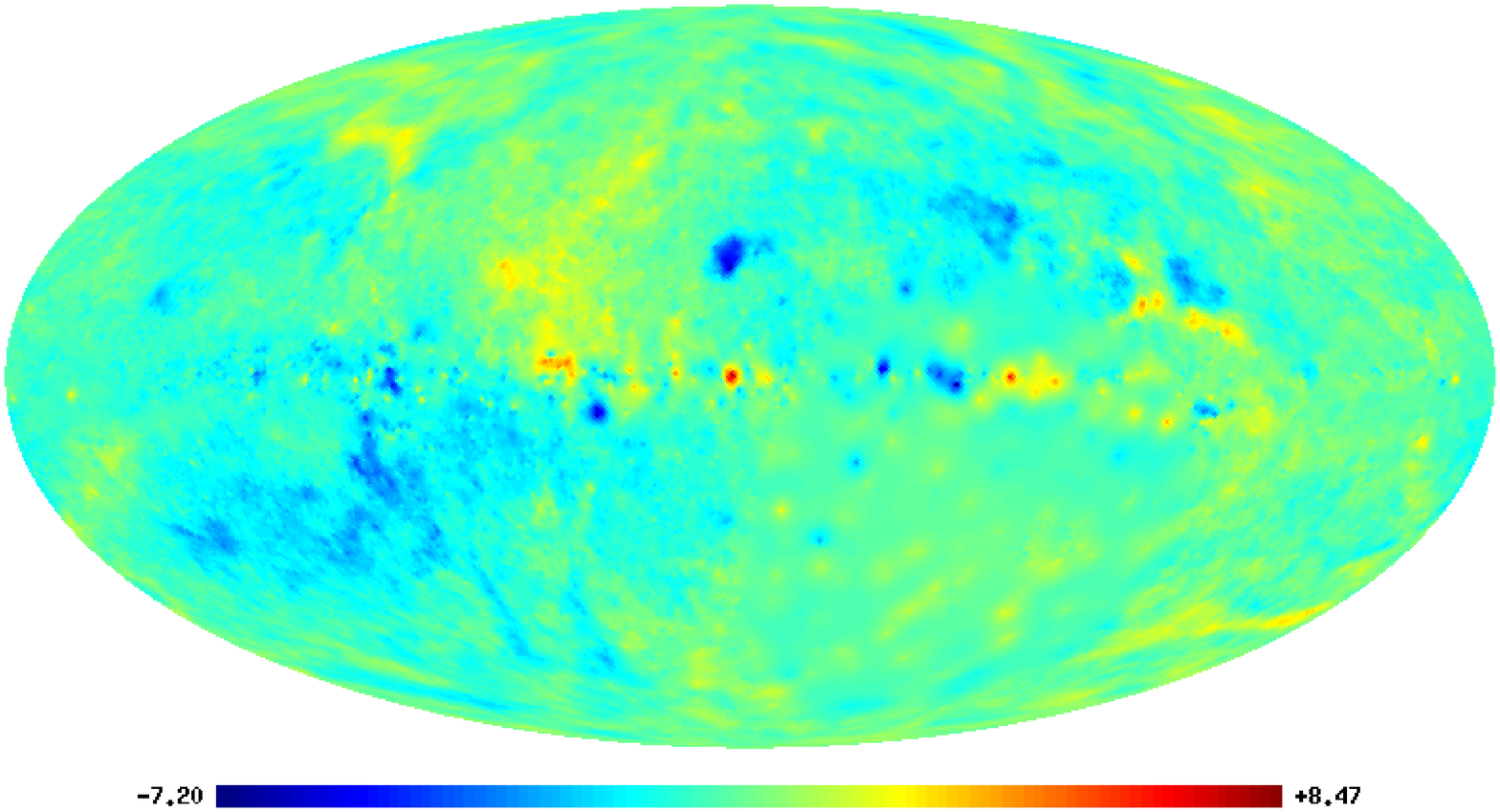}}
    \centerline{\includegraphics[width=1.0\linewidth]{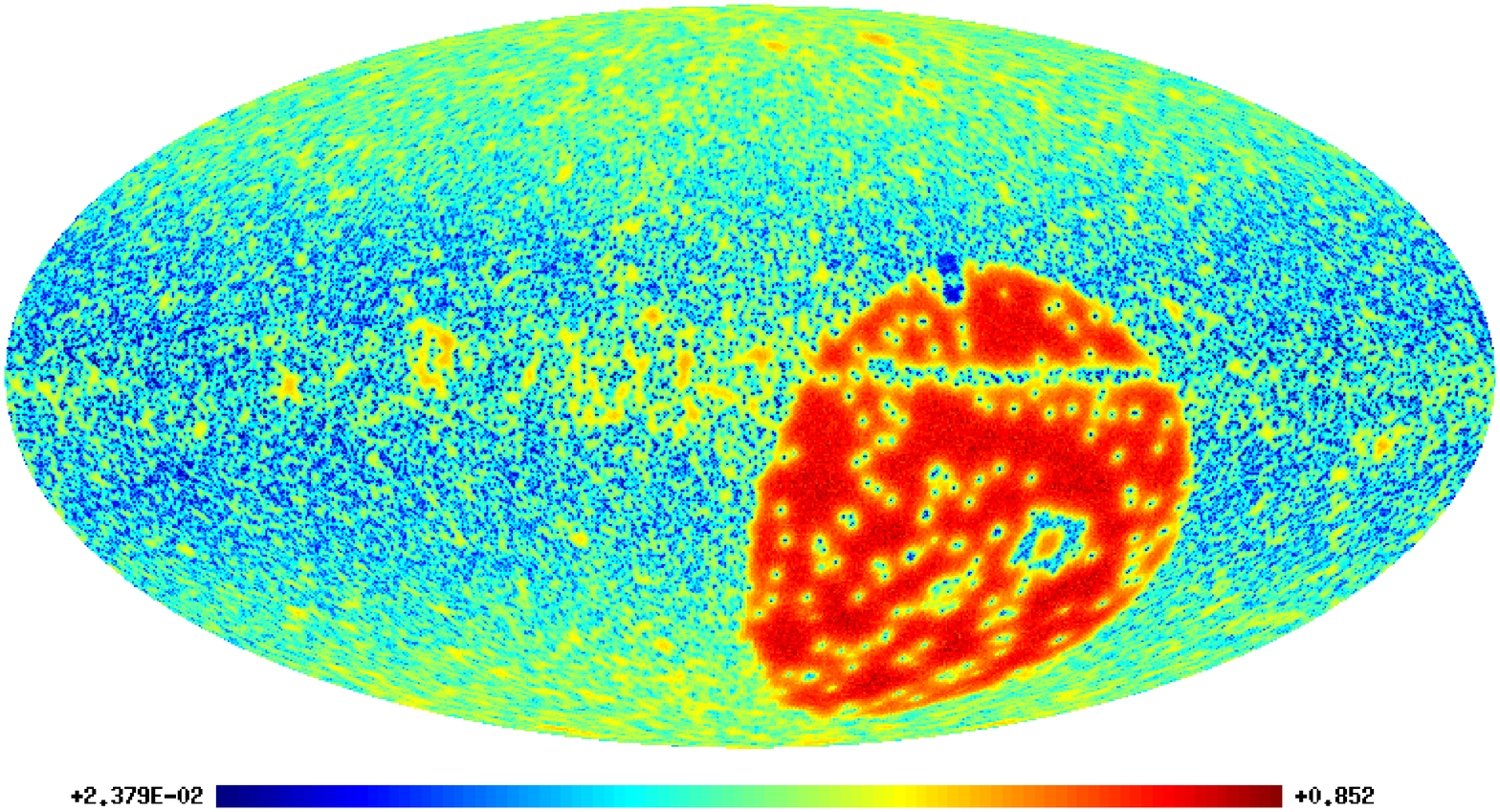}}
    \caption{The dimensionless signal map $s(l,b)$ (top) and its uncertainty (bottom).}
    \label{s-map}
  \end{center}
\end{figure}

\begin{figure}
  \begin{center}
    \centerline{\includegraphics[width=1.0\linewidth]{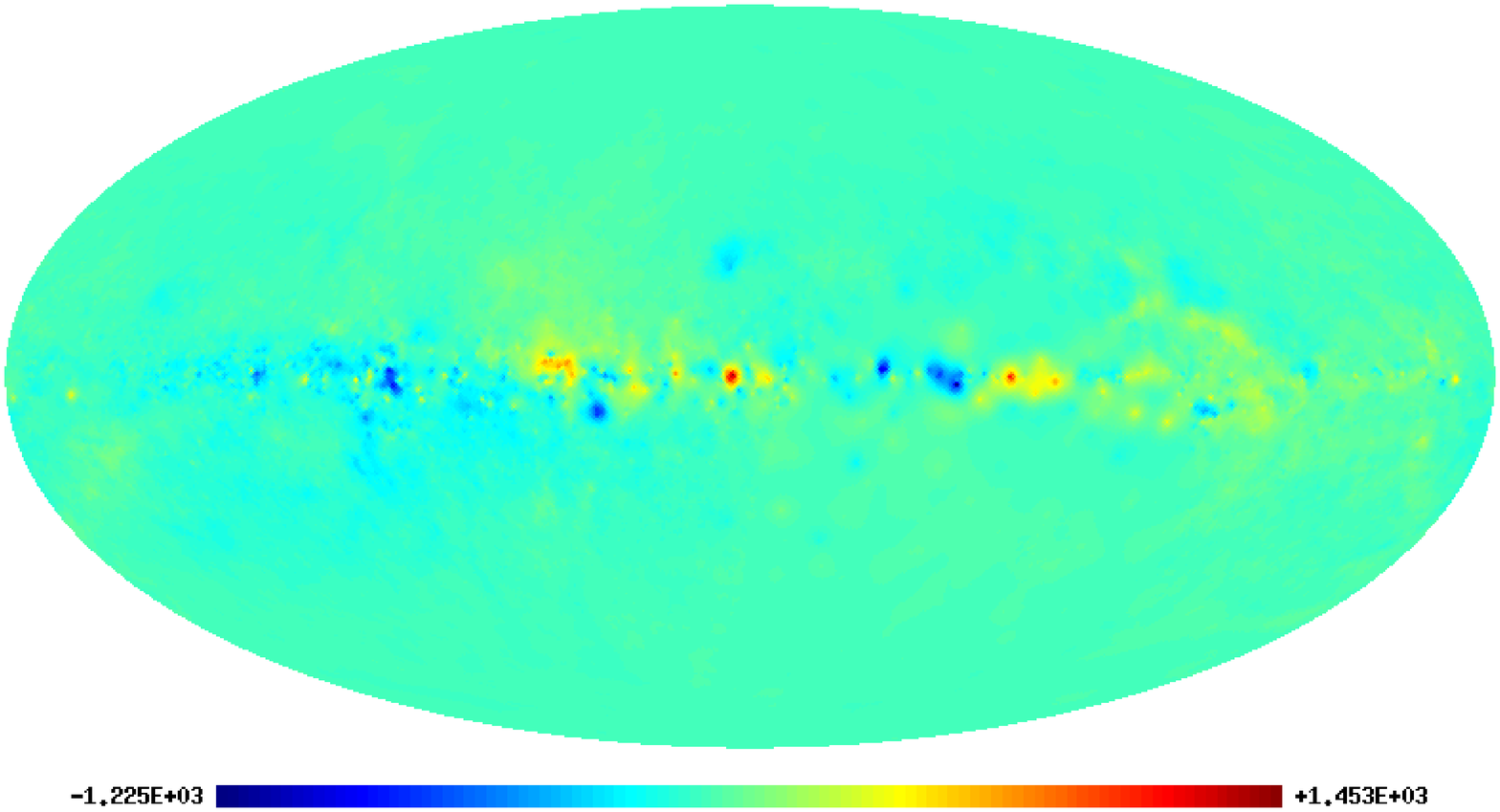}}
    \centerline{\includegraphics[width=1.0\linewidth]{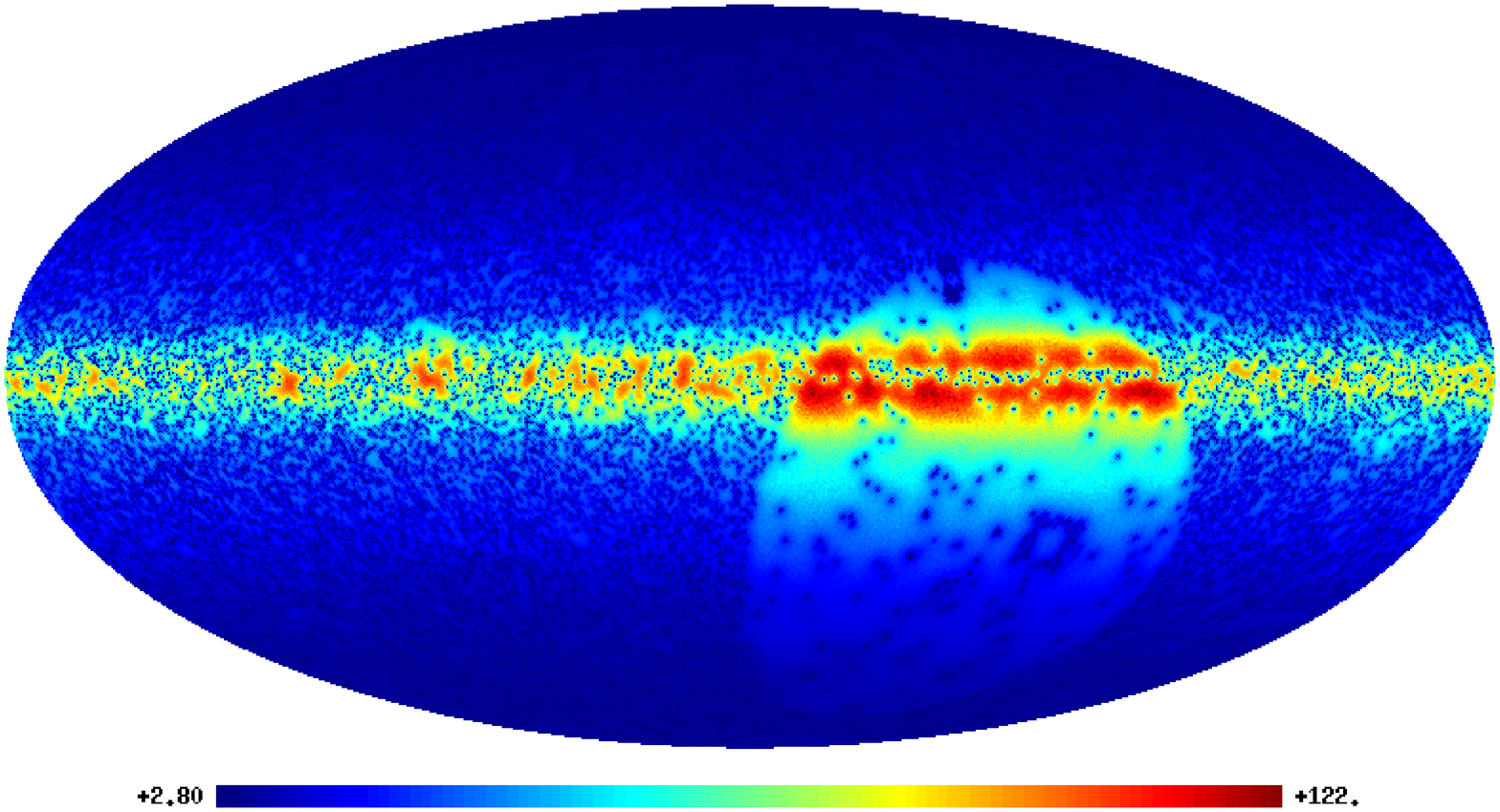}}
    \caption{The Galactic Faraday depth map $\phi(l,b)$ (top) and its uncertainty (bottom) in ${\rm rad/m^2}$.}
    \label{phi-map}
  \end{center}
\end{figure}

Faraday rotation measurements of the extragalactic radio sources provide the tracers of the Galactic magnetic field.
When the polarized radiation propagates through a magnetized plasma, the Faraday depth is given by Eq. \ref{eq4}.
In the recent work \citep{Faraday_depth}, the authors reconstructed the full-sky Galactic Faraday depth maps by using the point source Faraday depth measurements of the NRAO VLA SKY Survey (NVSS) catalog as well as other surveys. In the reconstruction, in order to divide out the most obvious large scale anisotropy introduced by the presence of the Galactic disk, the authors defined the dimensionless signal $s(l,b)$ as
\begin{eqnarray}\label{s-definition}
s(l,b)=\frac{\phi(l,b)}{p(b)},
\end{eqnarray}
where $\phi(l,b)$ is the potential Faraday depth map. The variance profile $p(b)$ is a function of Galactic latitude only. 
We show the reconstructed signal map $s(l,b)$ and the derived Faraday depth map $\phi(l,b)$ in Fig. \ref{s-map} and Fig. \ref{phi-map}, which have the HEALPix \citep{healpix} resolution parameter $N_{\rm side}=128$. In both figures, we see a large region with quite large uncertainties in the southern sky below the declination of $-40^{\circ}$, which is caused by the lack of data in this region due to the position of the observing telescope VLA (see \citep{Faraday-map0},\citep{Faraday_depth} and \citep{nvss} for the details). In our following analysis, we shall mask this region to reduce the uncertainties.

\subsection{Data of CMB temperature fluctuations}
\begin{figure}
  \begin{center}
    \centerline{\includegraphics[width=1.0\linewidth]{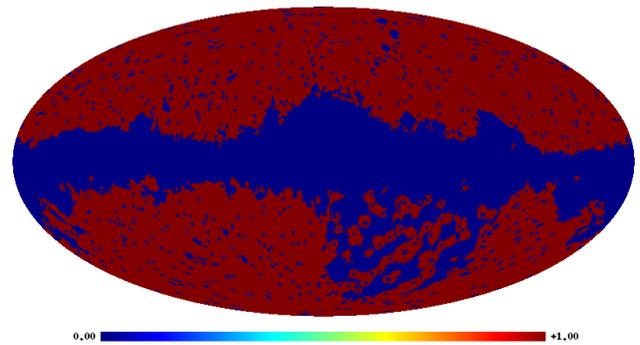}}
    \caption{The kq75p1 mask.}
    \label{kq75p1}
  \end{center}
\end{figure}

The WMAP instrument is composed of 10 differencing assemblies (DAs) spanning five frequencies from 23 to 94GHz \citep{wmapresults}. The internal linear combination (ILC) method has been used by the WMAP team to generate the WMAP ILC maps \citep{wmap3ytem},\citep{gold2011}. The 7-year ILC (ILC7) map is a weighted combination from all five original frequency bands, which are smoothed to a common resolution of one degree. In this paper, we shall consider the ILC7 map for the analysis.\\
Also, we will analyze the 7-year WMAP foreground reduced Q-band (Q7), V-band (V7) and W-band (W7) maps, as well as the co-added maps of these frequency bands. The linearly co-added maps are constructed using an inverse weight of pixel-noise variance. We adopt two maps with different combinations of frequency bands: V-band and W-band (written as `VW7') and Q-band, V-band and W-band (written as `QVW7'). Note that all these WMAP data have the same resolution parameter $N_{\rm side}=512$. In order to cross-correlate with Faraday depth maps, we degrade them to a lower resolution $N_{\rm side}=128$ for the analysis.\\
To exclude the effect of various contaminations, in the analysis we apply the KQ75p1 mask, which is a combination of KQ75y7 mask given by WMAP team and a mask which rejects the regions with large uncertainty in the Faraday signal map, i.e. $\sigma_{s}(l,b)>0.7$ (see the right panel in Fig. \ref{s-map}).  The mask is shown in Fig. \ref{kq75p1}.

\section{Regions of strongest correlation}
\begin{figure}
  \begin{center}
    \centerline{\includegraphics[width=1.0\linewidth]{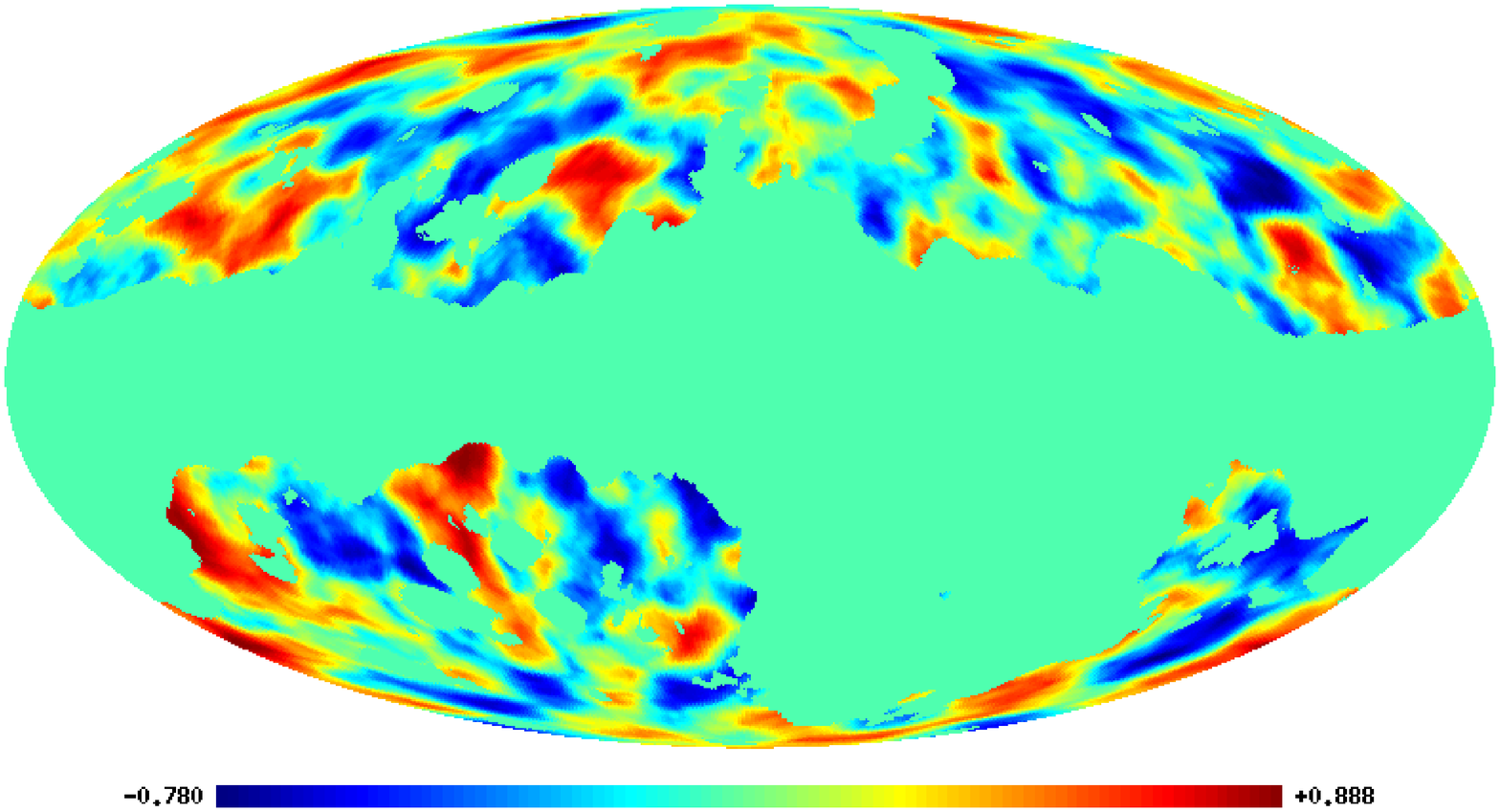}}
    \centerline{\includegraphics[width=1.0\linewidth]{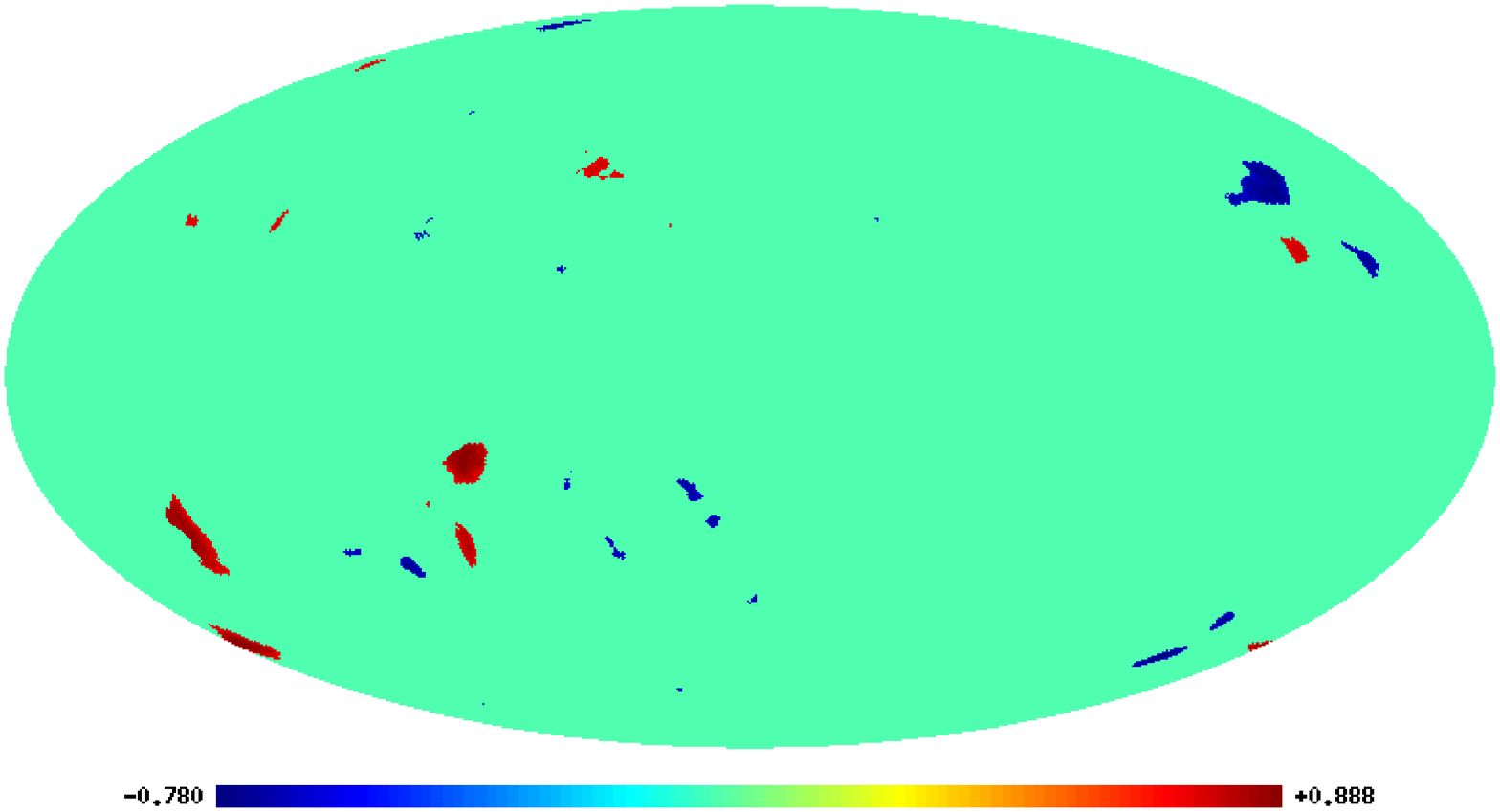}}
    \caption{Top: $r_s(l,b,R)$ with $R=5^{\circ}$ based on Faraday $\phi$-map and ILC7 map; Bottom: same as top one, 
but here we have set $r_s=0$ if $|r_s-\bar{r_s}|<2\sigma_{r_s}$.  }
    \label{ilc7-5deg}
  \end{center}
\end{figure}

\begin{figure}
  \begin{center}
    \centerline{\includegraphics[width=1.0\linewidth]{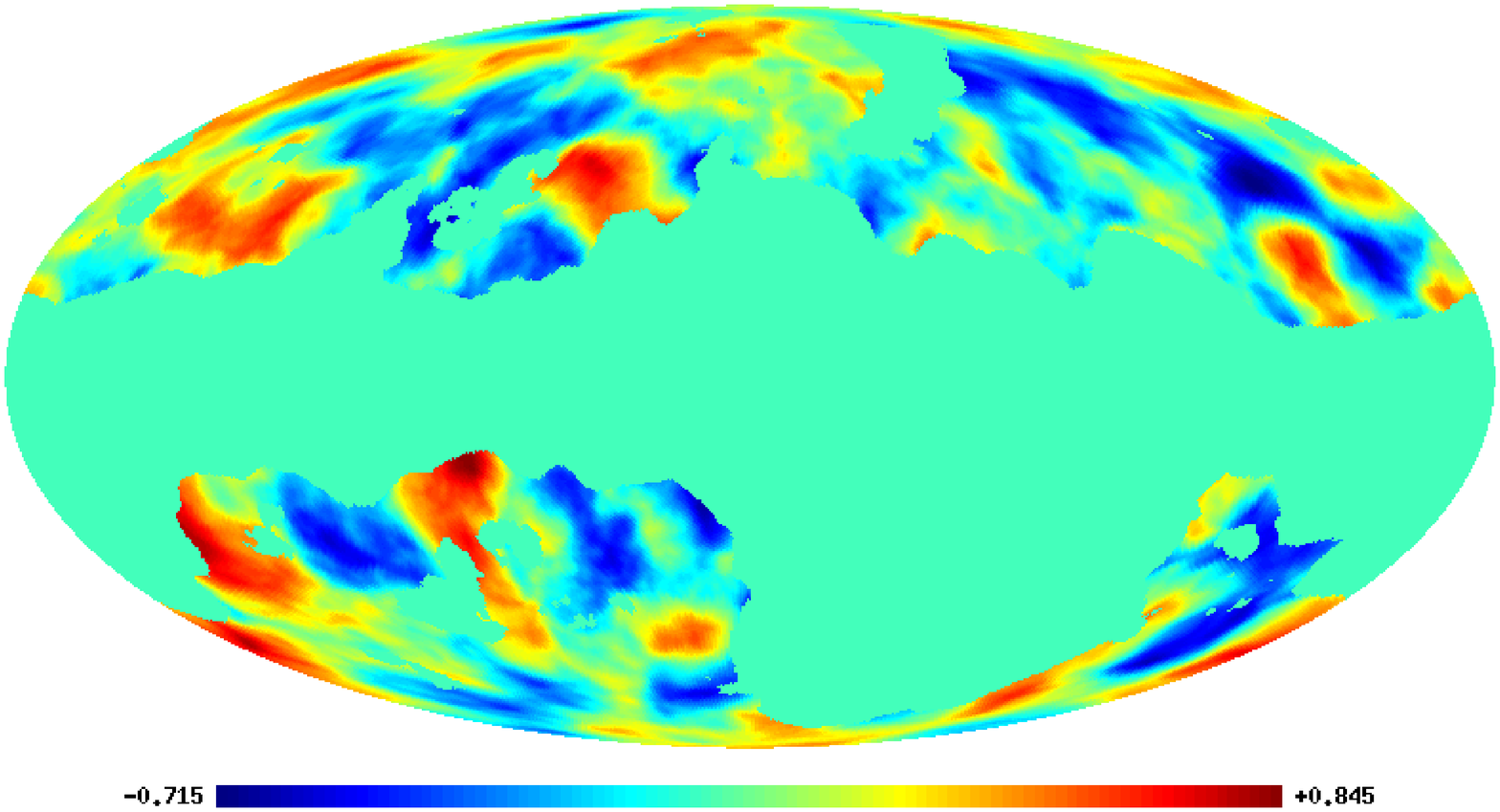}}
    \centerline{\includegraphics[width=1.0\linewidth]{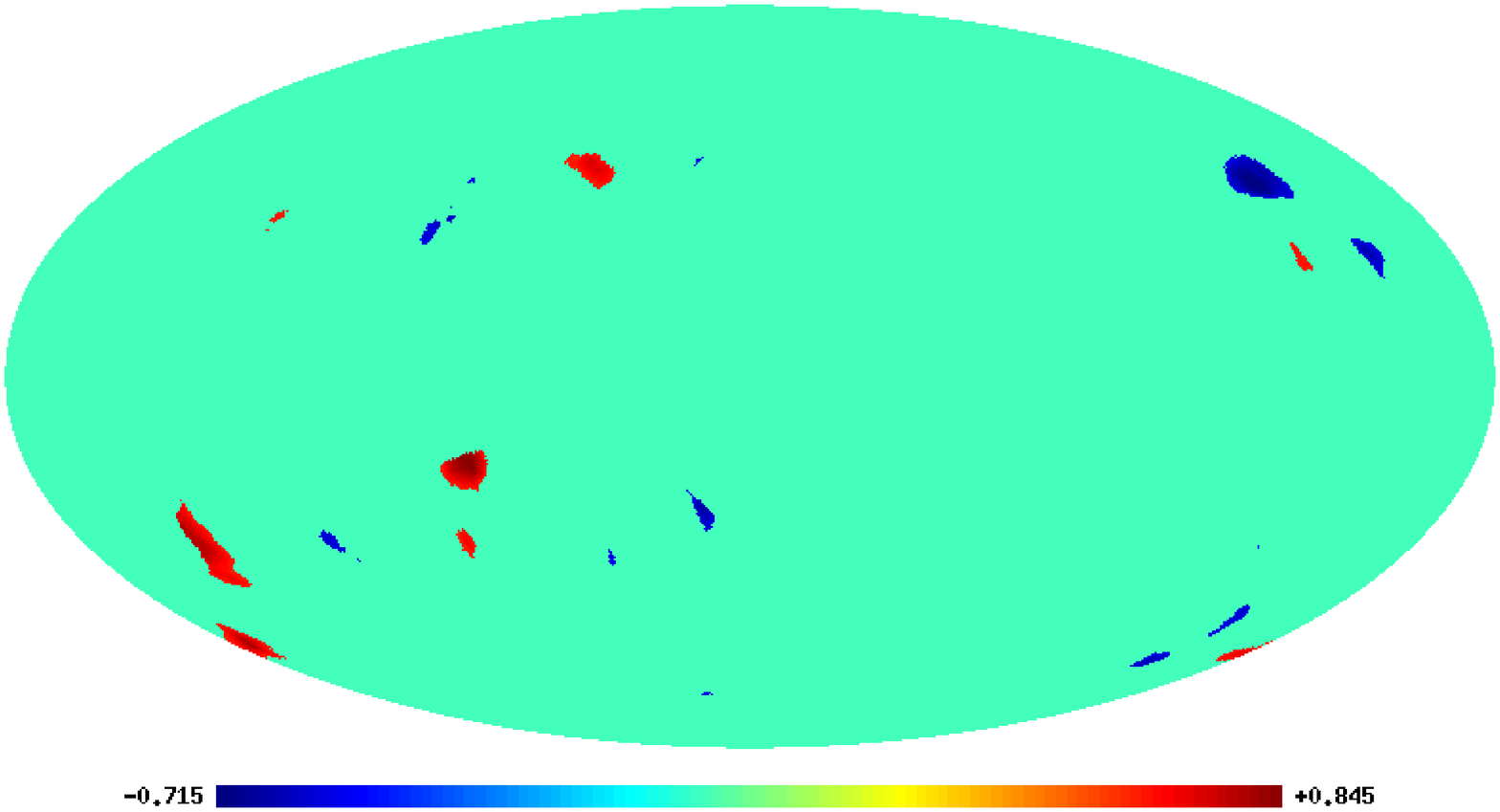}}
    \caption{Top: $r_s(l,b,R)$ with $R=7^{\circ}$ based on Faraday $\phi$-map and ILC7 map; Bottom: same as top one, 
but here we have set $r_s=0$ if $|r_s-\bar{r_s}|<2\sigma_{r_s}$.}
    \label{ilc7-7deg}
  \end{center}
\end{figure}

We now construct the $r_s(l,b,R)$ map by the following steps:\\
1) We degrade the CMB temperature anisotropy maps (ILC7, Q7, V7, W7, QVW7, VW7) to the low resolution $N_{\rm side}=128$.\\
2) For the given Faraday map ($\phi$-map or $s$-map) and the degraded CMB map, we mask them by kq75p1 mask.\\
3) Based on these two masked maps, we calculate the values of $\bar{x}$ and $\bar{y}$.\\
4) Choosing an $R$ value, and using Eq.(\ref{smap}), we construct $r_s(l,b,R)$ map.\\
Note that for every $C$, we only take into account the unmasked pixels. In the output $r_s(l,b,R)$ map, we set $r_s(l_i,b_i,R)=0$ if more than $10\%$ pixels are masked in the $C$ around $(l_i,b_i)$. \\
In Figs. \ref{ilc7-5deg} and \ref{ilc7-7deg} (upper panels), we plot the $r_s(l,b,R)$ maps with $R=5^{\circ}$ and $R=7^{\circ}$ where the Faraday $\phi$-map and the ILC7 CMB map are used. Also, we calculated the same correlation for the full sky map (masked by the KQ75p1 mask), resulting in an overall correlation of $0.0504$.\\
In order to investigate the Faraday-CMB cross correlation further, we define two statistics: the mean value $\bar{r_s}$ and the standard deviation $\sigma_{r_s}$, which are calculated from the unmasked $r_s(l,b,R)$ maps. Specifically, we calculate the standard deviation as
\begin{eqnarray}
\sigma_{r_s}=\sqrt{\frac{1}{N} \sum_{i}(x_{i} - \bar{x})^{2}},
\end{eqnarray}
where $x_{i}$ is the $i$'th pixel value out of $N$ in the region, and $\bar{x}$ is the average pixel value for the unmasked pixels. \\
Now, for each $r_s(l,b,R)$ map, we calculate the mean value $\bar{r_s}$ and standard deviation $\sigma_{r_s}$, and only show the areas with strongest correlation, i.e. $|r_s-\bar{r_s}| > 2\sigma_{r_s}$. The results are presented in Figs. \ref{ilc7-5deg} and \ref{ilc7-7deg} (bottom panels).  We have also checked that we obtain similar results, if we replace the Fararay $\phi$-map with the $s$-map, or replace the ILC7 map with the other CMB maps, or consider other cases with $R=3^{\circ}$ and $9^{\circ}$.

\begin{figure}
  \begin{center}
    \centerline{\includegraphics[width=0.80\linewidth]{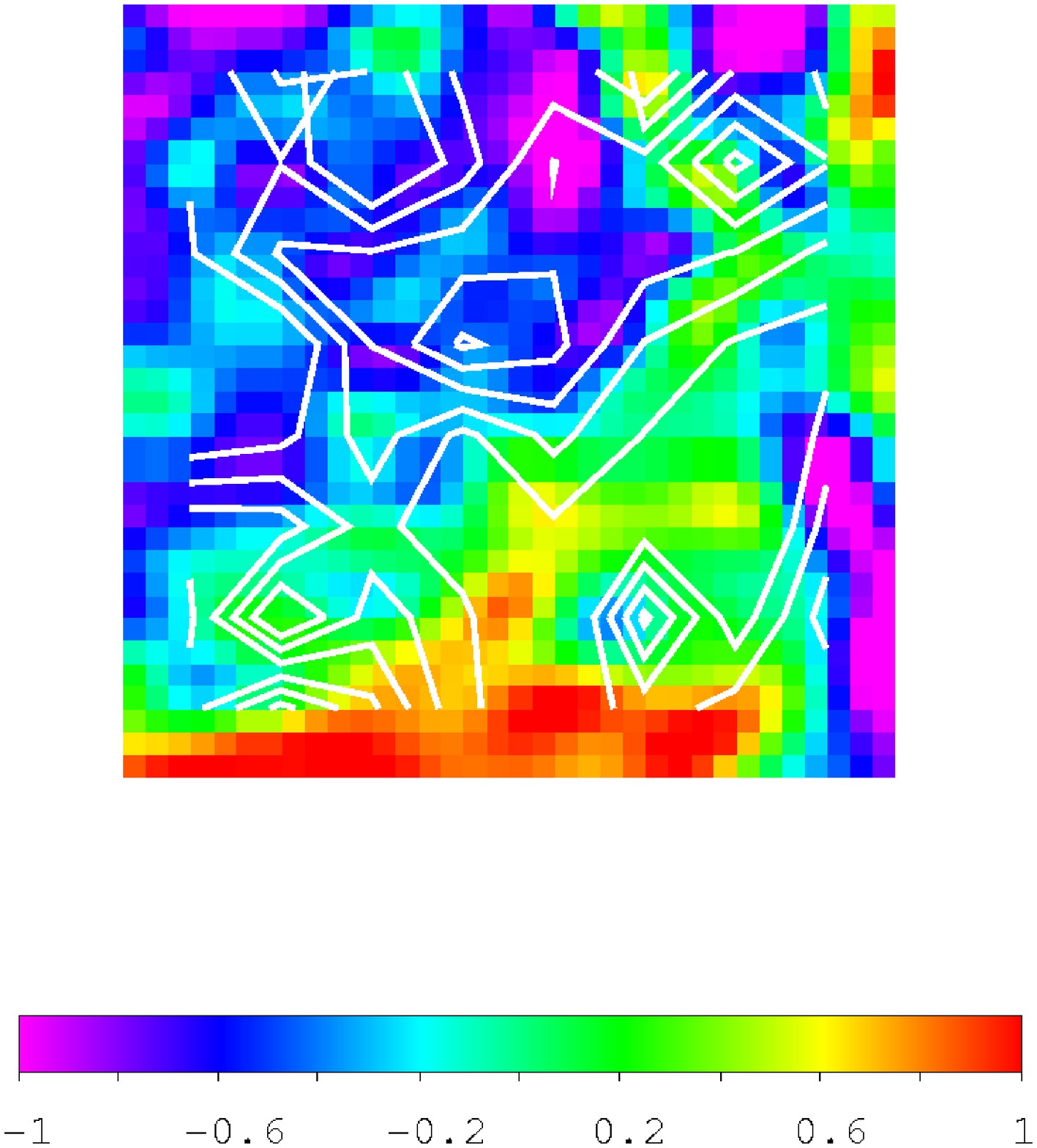}}
    \centerline{\includegraphics[width=0.80\linewidth]{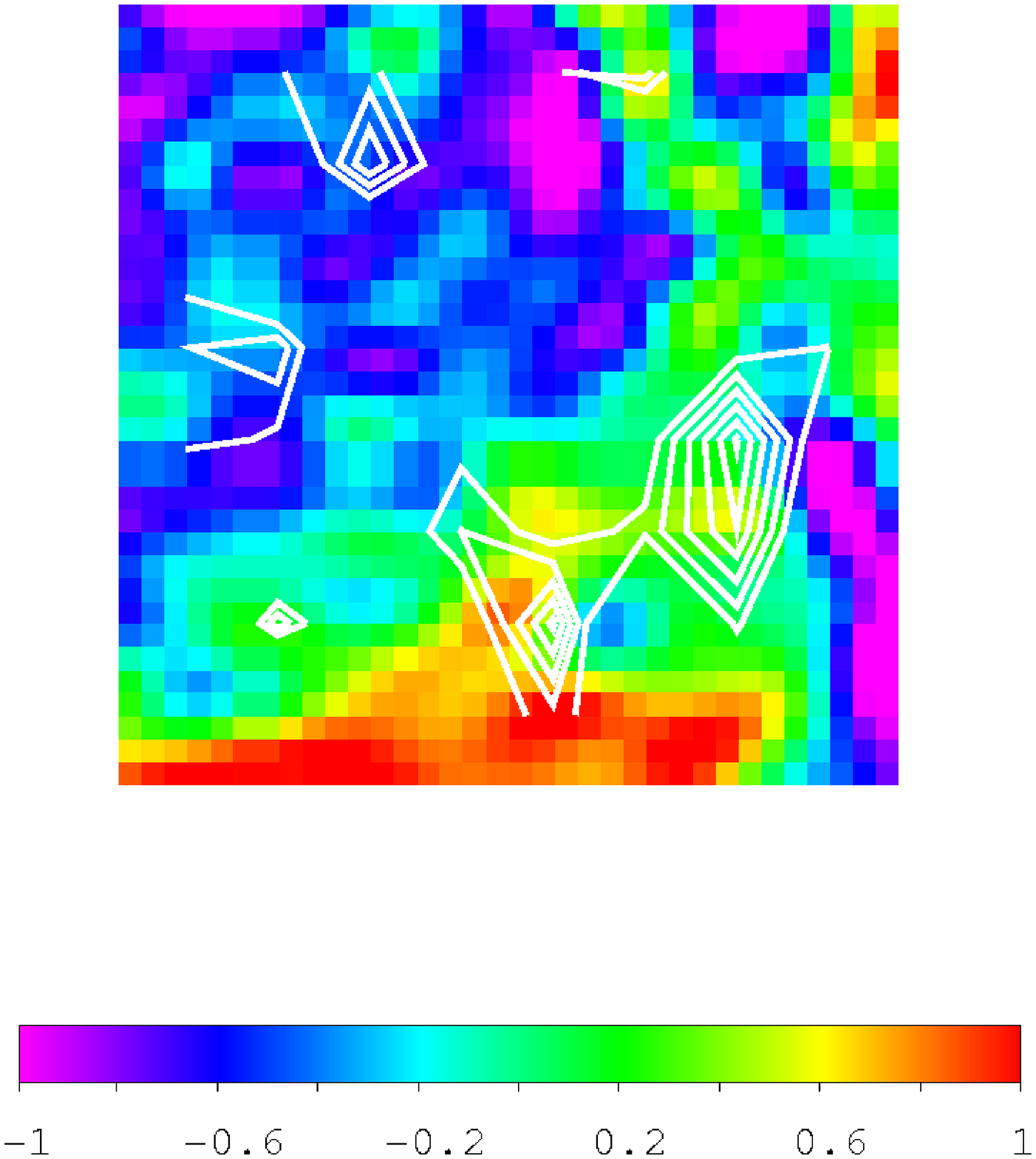}}
    \caption{Top. Isolines of the ILC 7 CMB temperature fluctuations on the Faraday rotation depth map, for the zone with Galactic coordinates $l=37^\circ,b=54^\circ$. The isolines start at $-0.025$ mK, and go down with a step of $-0.025$ mK. Bottom. The same, as the top, but for all levels from $0.0$mK and up, with a step $0.0125$ mK. Note the non-symmetric color scale, in which magenta indicate negative signal, while, starting from blue, the Faraday depth is positive.}
    \label{fig_FC1}
  \end{center}
\end{figure}

\begin{figure}
  \begin{center}
    \centerline{\includegraphics[width=0.80\linewidth]{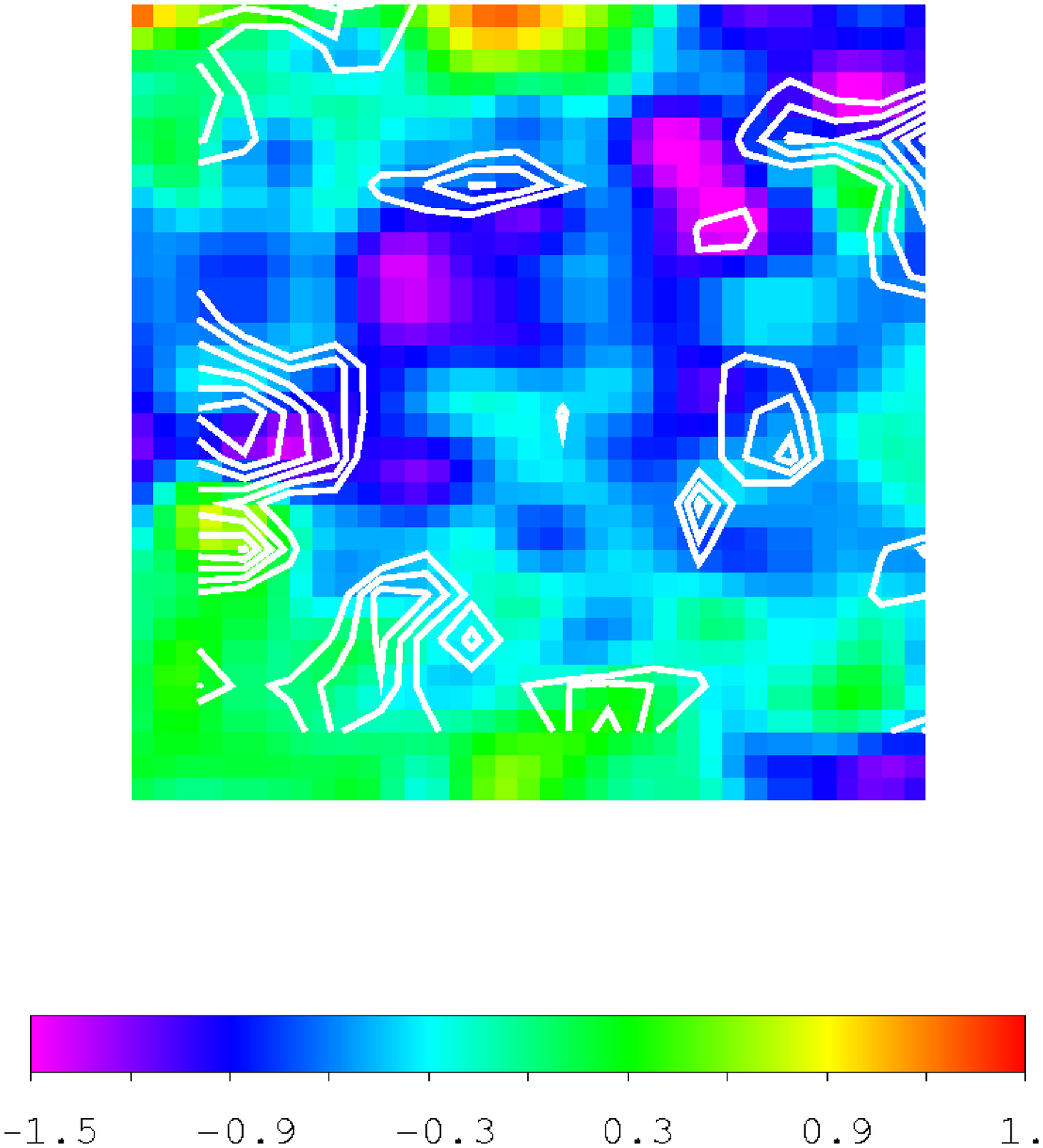}}
    \centerline{\includegraphics[width=0.80\linewidth]{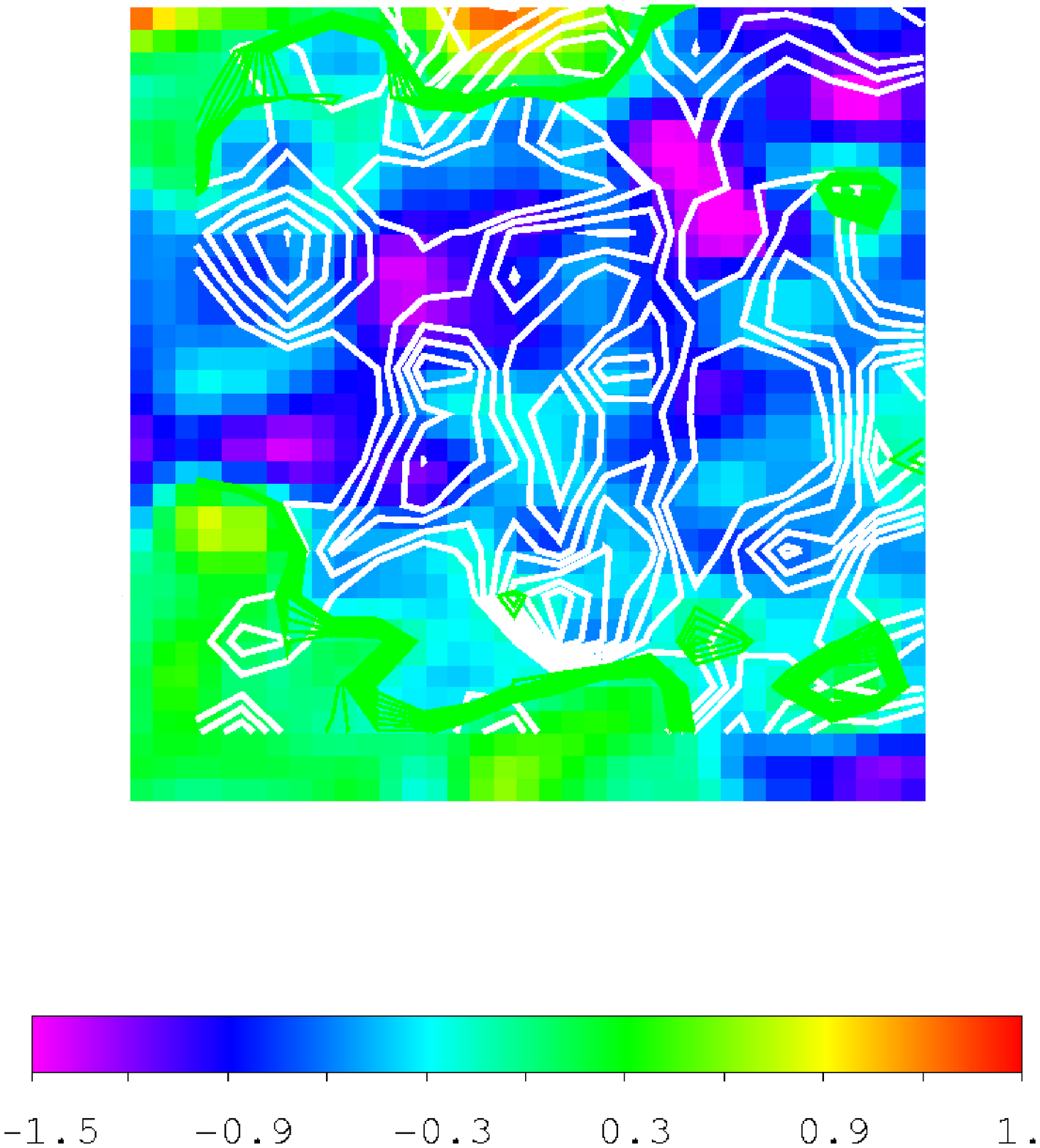}}
    \caption{Similar to Fig. \ref{fig_FC1}, but at Galactic coordinates $l=125^\circ,b=47^\circ$}
    \label{fig_FC2}
  \end{center}
\end{figure}

\begin{figure}
  \begin{center}
    \centerline{\includegraphics[width=0.80\linewidth]{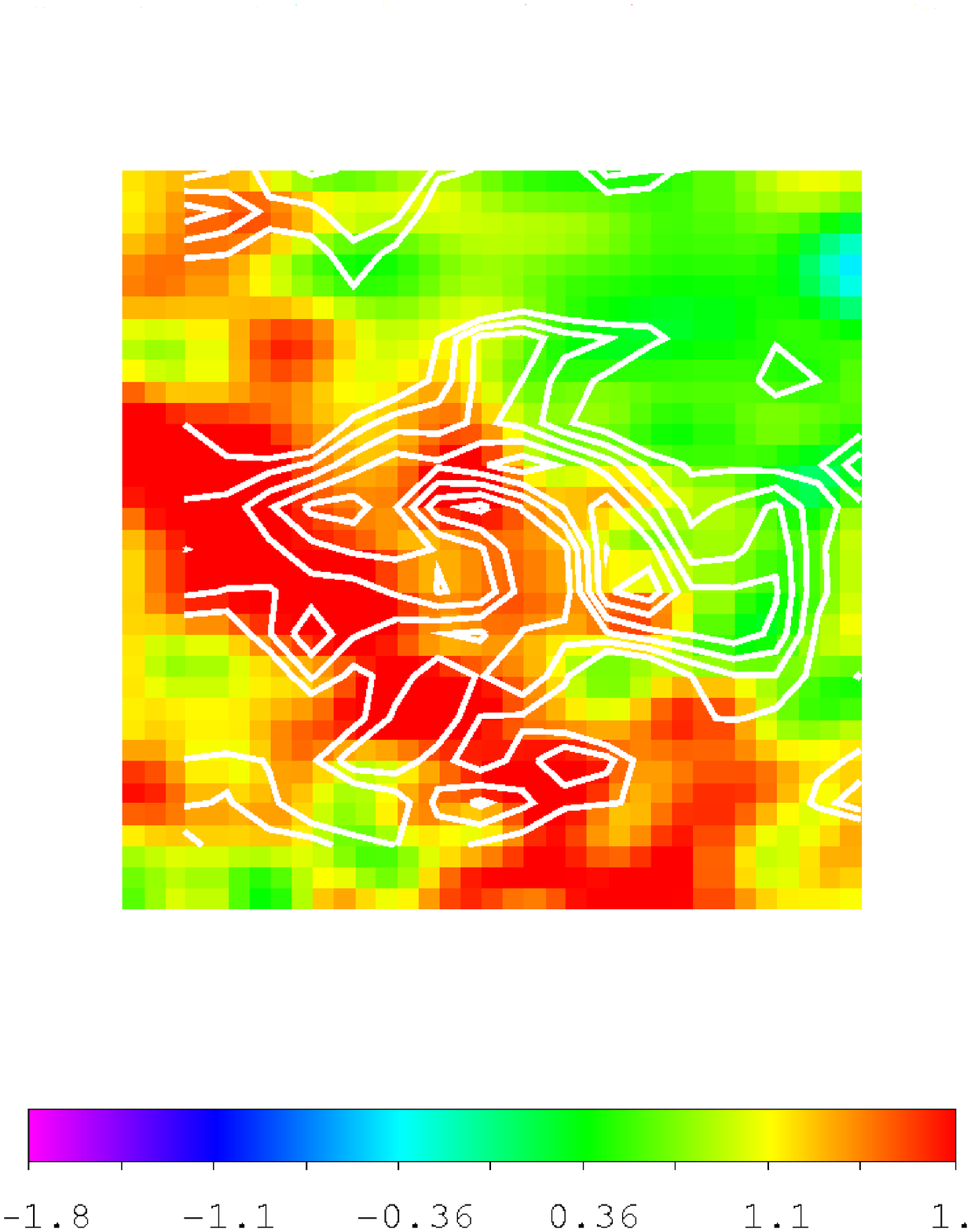}}
    \centerline{\includegraphics[width=0.80\linewidth]{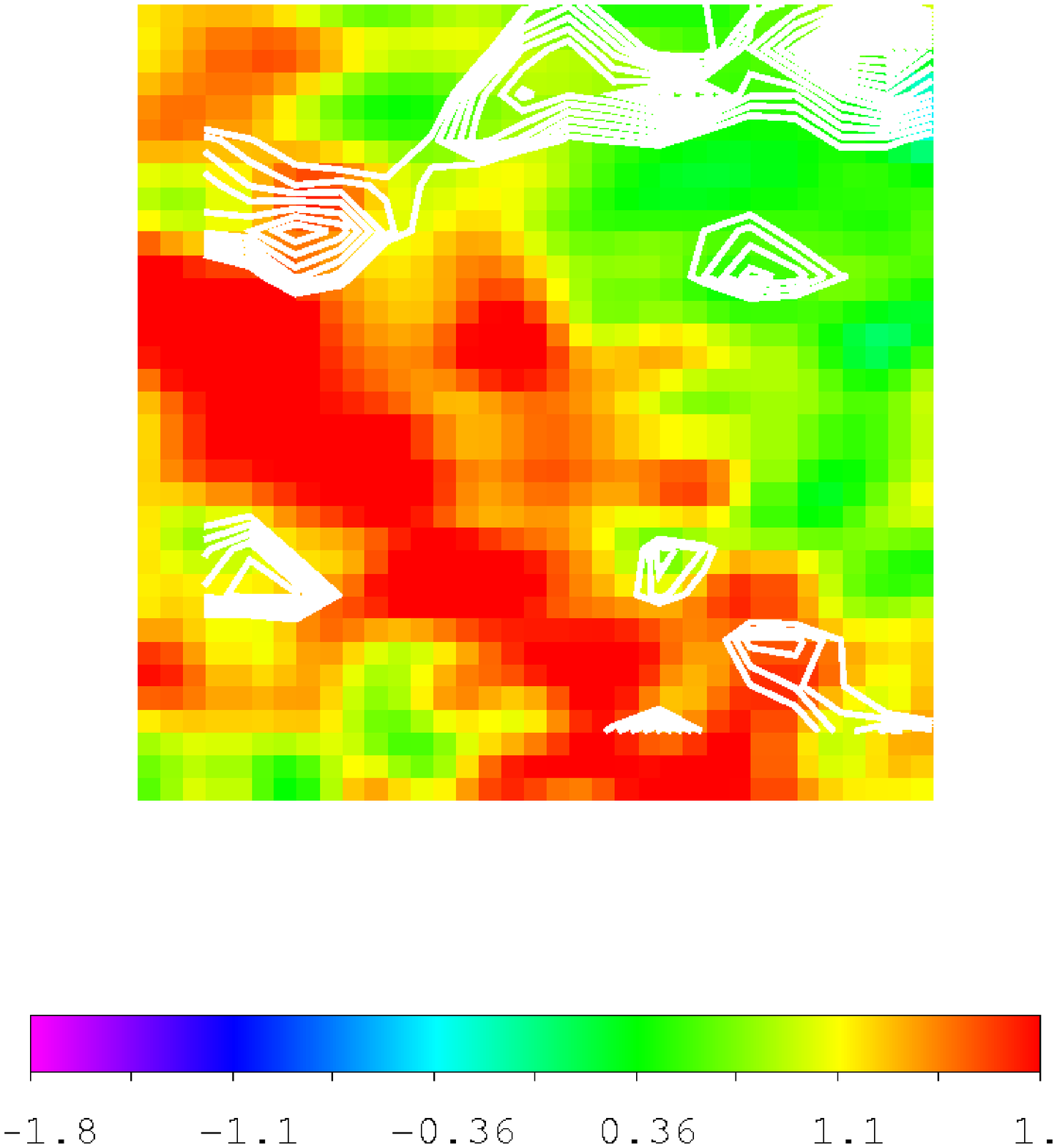}}
    \caption{Similar to Fig. \ref{fig_FC1}, but at Galactic coordinates $l=44^\circ,b=14^\circ$}
    \label{fig_FC3}
  \end{center}
\end{figure}

\begin{figure}
  \begin{center}
    \centerline{\includegraphics[width=0.80\linewidth]{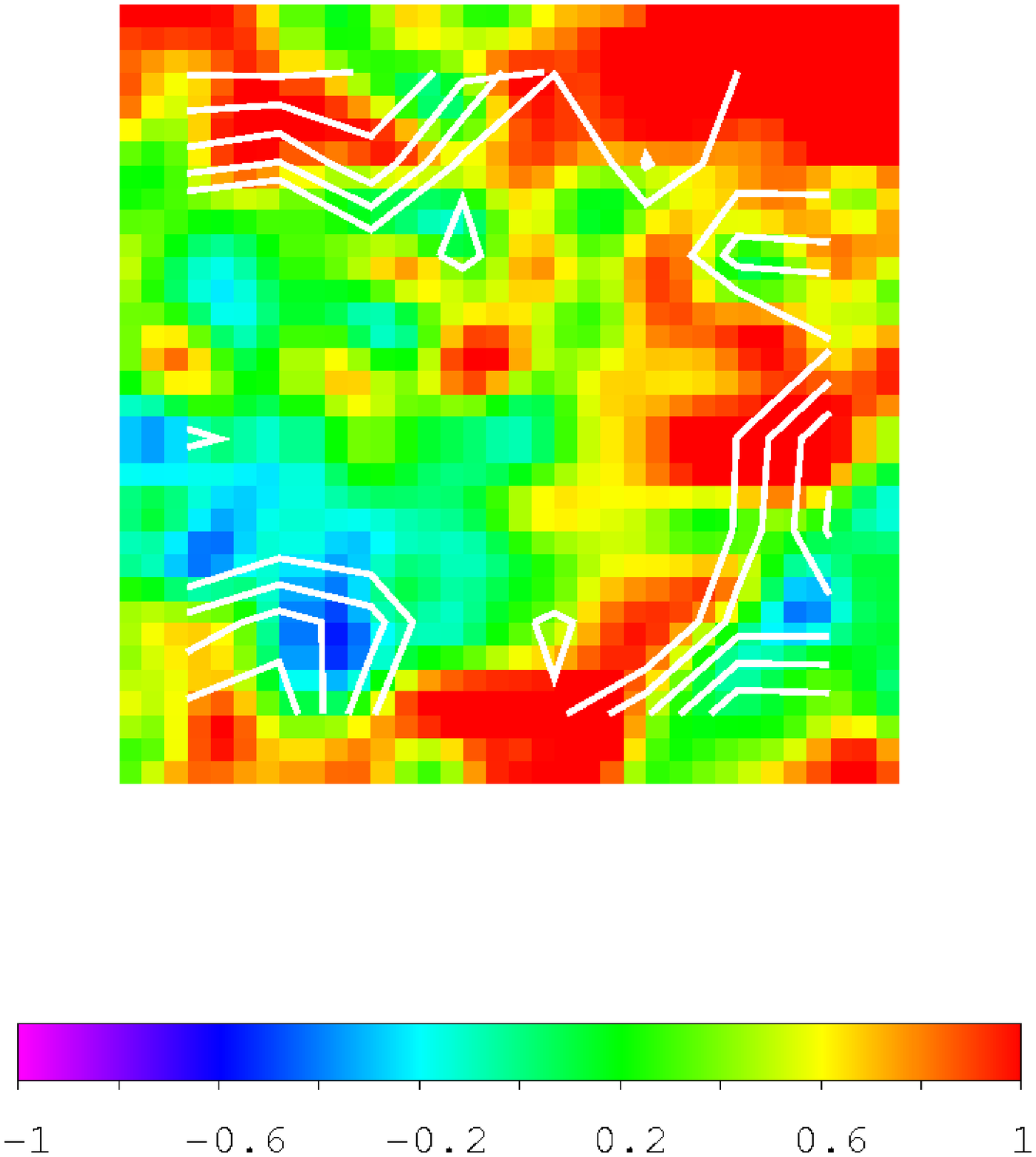}}
    \centerline{\includegraphics[width=0.80\linewidth]{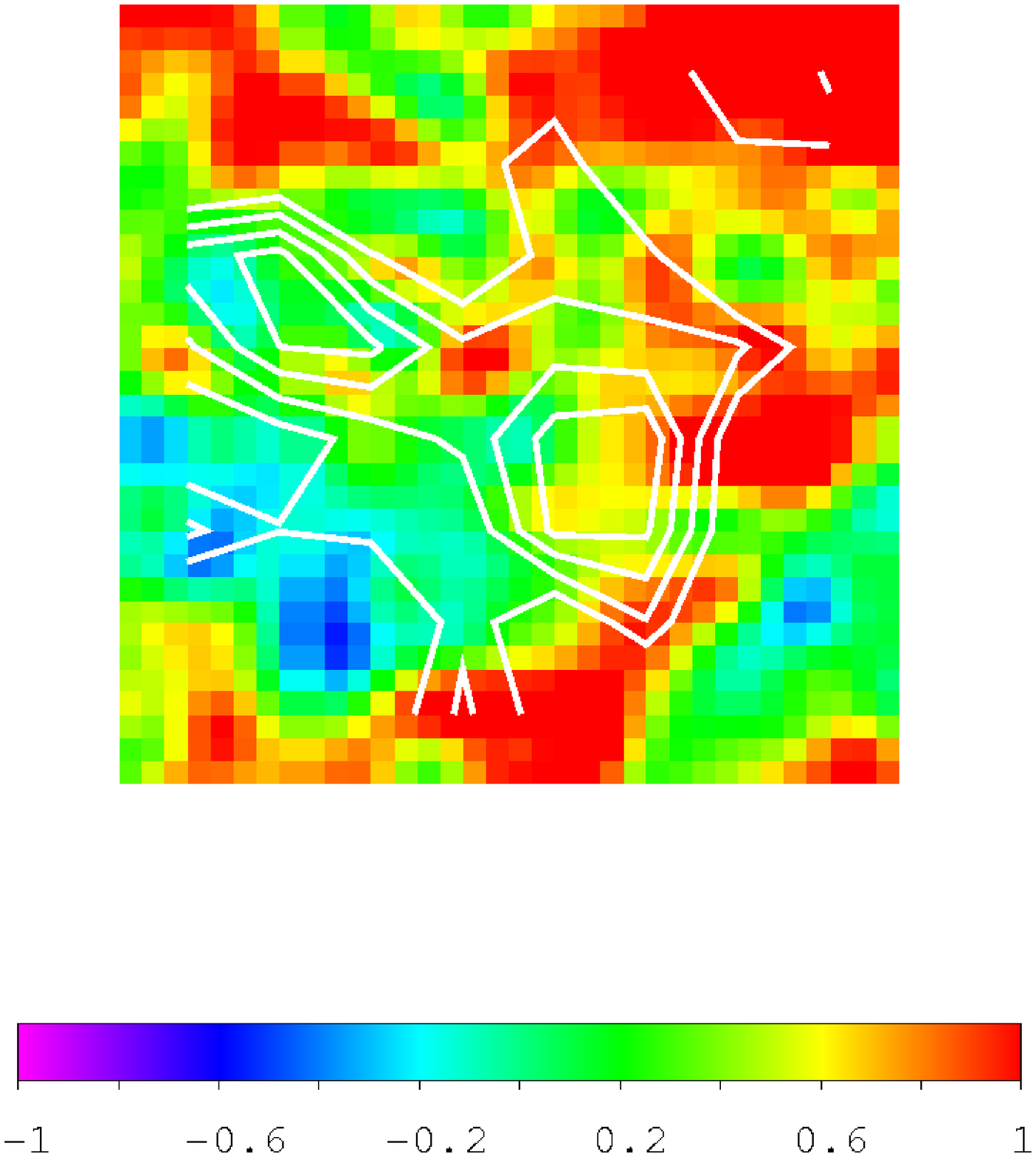}}
    \caption{Similar to Fig. \ref{fig_FC1}, but at Galactic coordinates $l=200^\circ,b=54^\circ$}
    \label{fig_FC4}
  \end{center}
\end{figure}

According to Fig.s \ref{ilc7-5deg} and \ref{ilc7-7deg} we have several zones of peculiar cross-correlations between the Faraday rotation map and the CMB temperature map. It is interesting to see, whether or not the extrema of the Faraday depth coincide with extreme temperatures in the CMB map. To investigate this, we present in Figs. \ref{fig_FC1}, \ref{fig_FC2}, \ref{fig_FC3} and \ref{fig_FC4} the CMB temperature isolines overlaid upon the Faraday depth map for 4 such areas. Clearly, we see several areas, where the concentration of isolines and the Faraday depth extrema overlap. This indicate a connection between the galactic magnetic field and the extreme temperatures of the CMB. 

\subsection{Comparing with simulations}
\begin{figure}
  \begin{center}
    \centerline{\includegraphics[width=1.10\linewidth]{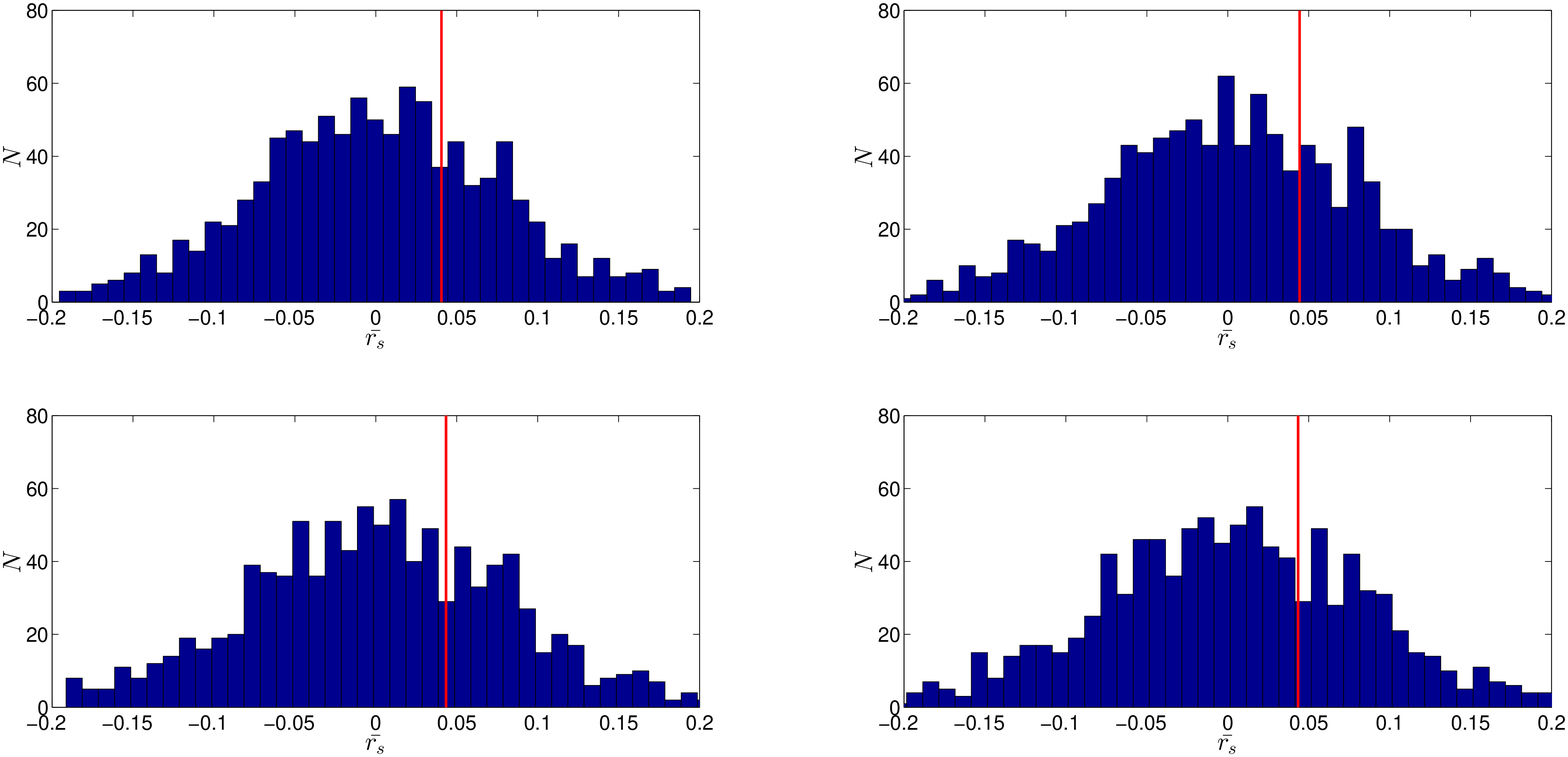}}
    \caption{The values of $\bar{r_s}$ for ILC7 and Faraday $\phi$-map (red line) compared with those for 1000 simulations. 
Upper left is for $R=3^{\circ}$, upper right is for $R=5^{\circ}$, lower left is for $R=7^{\circ}$, and lower right is for $R=9^{\circ}$.}
    \label{bar-ilc7}
  \end{center}
\end{figure}

\begin{figure}
  \begin{center}
    \centerline{\includegraphics[width=1.10\linewidth]{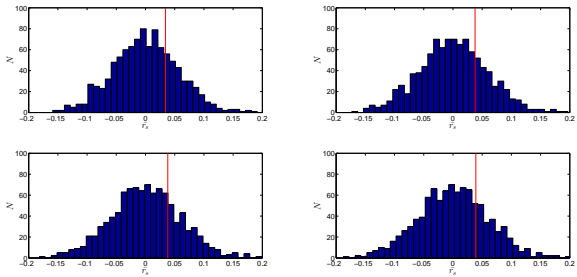}}
    \caption{Same as Fig. \ref{bar-ilc7} but the ILC7 map is replaced by the W7 map.} 
    \label{bar-w7}
  \end{center}
\end{figure}

\begin{figure}
  \begin{center}
    \centerline{\includegraphics[width=1.0\linewidth]{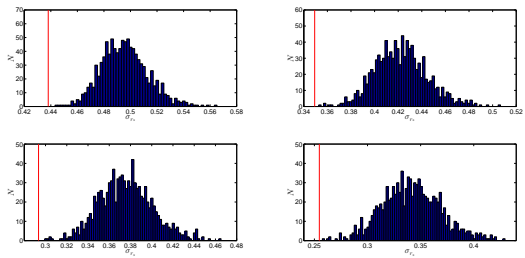}}
    \caption{The values of $\sigma_{r_s}$ for ILC7 (red line) comparing with those for 1000 simulations. 
Upper left is for $R=3^{\circ}$, upper right is for $R=5^{\circ}$, lower left is for $R=7^{\circ}$, and lower right is for $R=9^{\circ}$.}
    \label{sigma-ilc7}
  \end{center}
\end{figure}

\begin{figure}
  \begin{center}
    \centerline{\includegraphics[width=1.0\linewidth]{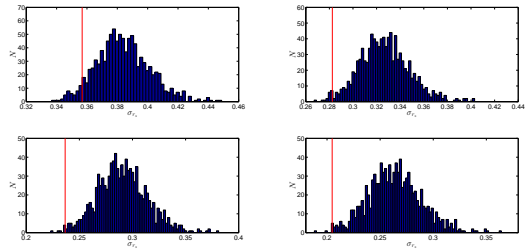}}
    \caption{Same with Fig. \ref{sigma-ilc7} but ILC7 map is replaced by W7 map.}
    \label{sigma-w7}
  \end{center}
\end{figure}

In order to study the global properties of the cross-correlation between Faraday rotation maps and the WMAP data, 
we shall compare the $r_s$ maps based on WMAP data with those of the random CMB simulations.\\
The cosmology in the random Gaussian simulation is based on the $\Lambda$CDM with the cosmological parameters from WMAP 7-year best-fit \citep{komatsu2011}: $\Omega_b h^2=0.02255$, $\Omega_c h^2=0.1126$, $\Omega_{\Lambda}=0.725$, $h=0.702$, $\tau=0.088$, $\Delta^2_{\mathcal R}=2.430\times10^{-9}$, $n_s=0.968$. First, we simulate the CMB maps for each frequency channel by considering the WMAP beam resolution and instrument noise for each channel, and then co-add them with inverse weight of the full-sky averaged pixel-noise variance in each frequency to get the simulated QVW7 and VW7 maps. To simulate the ILC7 maps, we consider the $1^{\circ}$ smoothing resolution and ignore the noises. In the simulation, we assumed that the temperature fluctuations and instrument noise follows the Gaussian distribution, and do not consider any effect due to the residual foreground contaminations. We have analyzed one thousand WMAP simulated data sets in the same way as the real data set.\\
Since the physical mechanism for the cross-correlation between Faraday rotation maps and the CMB temperature anisotropy maps is quite complicated, which depends on both the magnetic distribution and the thermal electron distribution, we cannot expect to know how the location and amount of highly correlated areas in the WMAP deviate from those of random Gaussian simulations.\\
We compare the $\bar{r_s}$ and $\sigma_{r_s}$ from the WMAP data with those of 1000 random simulations, and calculate the following probability functions
\begin{eqnarray}
\alpha={\rm Prob}(|{\bar{r_s}}^0|>|\bar{r_s}|),~~\beta={\rm Prob}(\sigma^0_{r_s}>\sigma_{r_s}),
\end{eqnarray}
where ${\bar{r_s}}^0$ and $\sigma^0_{r_s}$ are the values for the WMAP data, while $\bar{r_s}$ and $\sigma_{r_s}$ are those for 1000 CMB realizations. If there is no correlation between Faraday rotation maps and the WMAP data, we expect that the values of $\alpha$ and $\beta$ should be close to $0.5$. A larger deviation from $0.5$ indicates a stronger correlation. \\
We calculate $\alpha$ and $\beta$ values for all the combinations of the Faraday maps ($\phi$-map, $s$-map) and CMB maps (ILC7, QVW7, VW7, W7, V7, W7), by considering the cases with $R=3^{\circ}$, $5^{\circ}$, $7^{\circ}$, $9^{\circ}$. We find that for the statistic ${\bar{r_s}}$, the deviations from the random simulations are quite small, i.e. $\alpha\in[0.66,~0.74]$ is satisfied for all cases. These are clearly shown in Figs. \ref{bar-ilc7} and \ref{bar-w7}.\\
However, we find that the values $\sigma_{r_s}$ of WMAP data are quite smaller than those of simulations, i.e. the $\beta$ parameters are dramatically small for all the cases (see Tables \ref{beta-phi} and \ref{beta-s} and Figs. \ref{sigma-ilc7} and \ref{sigma-w7}). For instance, for all the cases related to ILC7 CMB map and Q7 map we always have $\beta \le 0.001$. So we conclude that the cross-correlations for these WMAP data deviate from those of random realizations at $99.9\%$ confidence level. From these Tables, we also find that the deviation from random realizations becomes smaller for the higher frequency bands. For the cases related to CMB W7 data, $\beta$ becomes $\sim 0.05$ for $R=3^{\circ}$ and $\sim 0.005$ for $R=9^{\circ}$. Even so, we also find that the cross-correlations for the WMAP data deviate from those of random realizations at more than $94.5\%$ confidence level. The dependence on the CMB frequency band can be explained as follows: in the higher frequency bands, the contaminations of Galactic magnetic fields and/or the thermal electrons (such as the residual synchrotron emissions) are smaller than those in the lower frequency bands.

\begin{table}
\begin{center}
\caption{Results of $\sigma_{r_s}$ and $\beta$, at $R=3^{\circ},5^{\circ},7^{\circ}$ and $9^{\circ}$, for Faraday $\phi$-map and various CMB data.}
\begin{tabular}{| c | r | r | r | r | r | r |}
\hline
   & ILC7 & QVW7 & VW7 & Q7 & V7 & W7\\
\hline
$\sigma^0_{r_s}$ ($3^{\circ}$)  & 0.438 & 0.372 & 0.362 & 0.379 & 0.360 & 0.357  \\
$\beta$ ($3^{\circ}$)           & 0.000  & 0.008  & 0.020  & 0.001  & 0.004  & 0.050   \\
\hline
$\sigma^0_{r_s}$ ($5^{\circ}$)  & 0.349 & 0.294 & 0.287 & 0.299 & 0.285 & 0.283  \\
$\beta$ ($5^{\circ}$)           & 0.000  & 0.002  & 0.003  & 0.000  & 0.001  & 0.016   \\
\hline
$\sigma^0_{r_s}$ ($7^{\circ}$)  & 0.294 & 0.246 & 0.240 & 0.251 & 0.239 & 0.237  \\
$\beta$ ($7^{\circ}$)           & 0.000  & 0.001  & 0.002  & 0.000  & 0.001  & 0.005   \\
\hline
$\sigma^0_{r_s}$ ($9^{\circ}$)  & 0.255 & 0.213 & 0.208 & 0.217 & 0.206 & 0.205  \\
$\beta$ ($9^{\circ}$)           & 0.000  & 0.001  & 0.002  & 0.001  & 0.001  & 0.005   \\
\hline
\end{tabular}
\end{center}
\label{beta-phi}
\end{table}

\begin{table}
\begin{center}
\caption{Results of $\sigma_{r_s}$ and $\beta$, at $R=3^{\circ},5^{\circ},7^{\circ}$ and $9^{\circ}$, for Faraday $s$-map and various CMB data.}
\begin{tabular}{| c | r | r | r | r | r | r |}
\hline
   & ILC7 & QVW7 & VW7 & Q7 & V7 & W7\\
\hline
$\sigma^0_{r_s}$ ($3^{\circ}$)  & 0.435 & 0.369 & 0.359 & 0.375 & 0.358 & 0.354  \\
$\beta$ ($3^{\circ}$)           & 0.000  & 0.009  & 0.020  & 0.000  & 0.006  & 0.054   \\
\hline
$\sigma^0_{r_s}$ ($5^{\circ}$)  & 0.344 & 0.290 & 0.282 & 0.295 & 0.281 & 0.278  \\
$\beta$ ($5^{\circ}$)           & 0.000  & 0.001  & 0.003  & 0.000  & 0.001  & 0.010   \\
\hline
$\sigma^0_{r_s}$ ($7^{\circ}$)  & 0.289 & 0.243 & 0.236 & 0.247 & 0.235 & 0.233  \\
$\beta$ ($7^{\circ}$)           & 0.000  & 0.001  & 0.001  & 0.000  & 0.001  & 0.004   \\
\hline
$\sigma^0_{r_s}$ ($9^{\circ}$)  & 0.250 & 0.209 & 0.203 & 0.213 & 0.202 & 0.201  \\
$\beta$ ($9^{\circ}$)           & 0.000  & 0.001  & 0.001  & 0.001  & 0.001  & 0.006   \\
\hline
\end{tabular}
\end{center}
\label{beta-s}
\end{table}

\section{Correlations at the Cold Spot}
\begin{figure}
  \begin{center}
    \centerline{\includegraphics[width=1.10\linewidth]{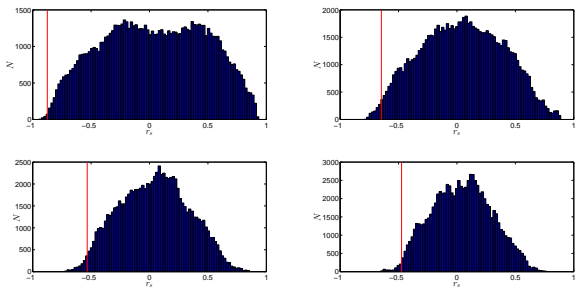}}
    \caption{Histogram for the $\phi$-map and ILC7 based $r_s(l,b,R)$ maps with $R=3^{\circ}$ (upper left), $R=5^{\circ}$ (upper right), $R=7^{\circ}$ (lower left), $R=9^{\circ}$ (lower right). In all panels, the red lines indicate the $r_s$ values around the CMB Cold Spot.}
    \label{his-ilc7}
  \end{center}
\end{figure}

\begin{figure}
  \begin{center}
    \centerline{\includegraphics[width=1.10\linewidth]{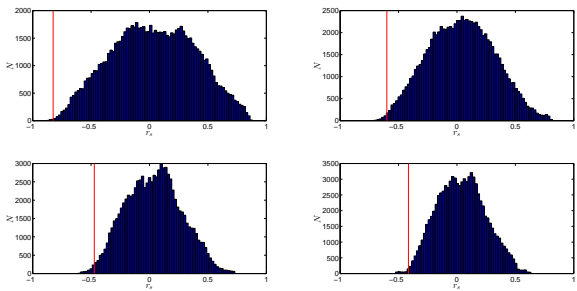}}
    \caption{Same as Fig. \ref{his-ilc7}, but for $r_s(l,b,R)$ based on $\phi$-map and CMB W7 map. }
    \label{his-w7}
  \end{center}
\end{figure}

It is most interesting, that from Figs. \ref{ilc7-5deg} and \ref{ilc7-7deg}, we find a strong negative correlation around the CMB Cold Spot at ($l=209^{\circ}$, $b=-57^{\circ}$) \citep{wmap7ylowl}.\\
Originally, the Cold Spot (CS) was detected in a wavelet analysis \citep{vielva2004}, \citep{cruz2005}, \citep{cjt} of the first-year data release from WMAP. It is apparently inconsistent with the assumption of statistically homogeneous Gaussian fluctuations, and its existence has later been confirmed \citep{cruz2006}, \citep{cruz2007}. Previously, \citep{rudnick} have analyzed the NVSS maps and have discovered a decrease in the space density of radio sources in the CS region. An independent study showed that the CS, may be a simple statistical deviation due to systematic effects \citep{smith_huterer}, or peculiarities of the low multipole tail of the CMB map at $2\le l \le 10$ \citep{cold_spot_env}. There are a few models of the primordial (cosmological) origin of the CS, based on textures \citep{Turok}, large voids \citep{Spergel} etc.\\
In order to quantify our findings regarding the CS, we investigate what the $r_s$-values is for the CS area, compared to the rest of the sky. In Figs. \ref{his-ilc7} and \ref{his-w7}, we present histograms of the masked $r_s(l,b,R)$ maps for four different radii ($R=3^{\circ}$, $R=5^{\circ}$, $R=7^{\circ}$ and $R=9^{\circ}$) for the ILC and W7 cases. The red line in the histograms corresponds to the value of $r_s$ around the CS. All histograms show that $r_s$ at the CS is smaller than mean value for the map at more than $2\sigma$ level. Clearly, the region around the CS show an anomalously strong anticorrelation, compared with the rest of the map.

\begin{figure}
  \begin{center}
    \centerline{\includegraphics[width=0.50\linewidth]{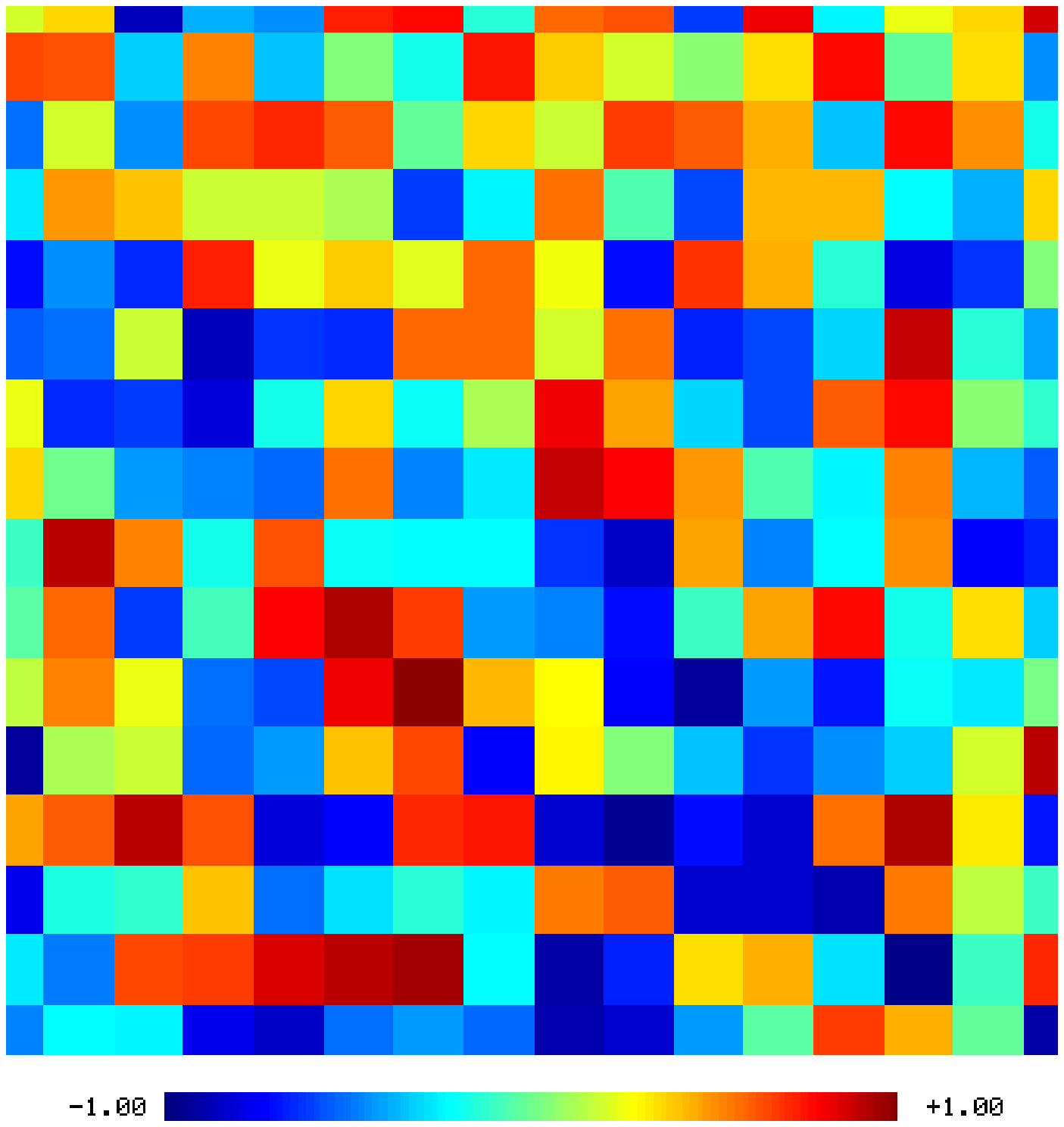} 
                \includegraphics[width=0.50\linewidth]{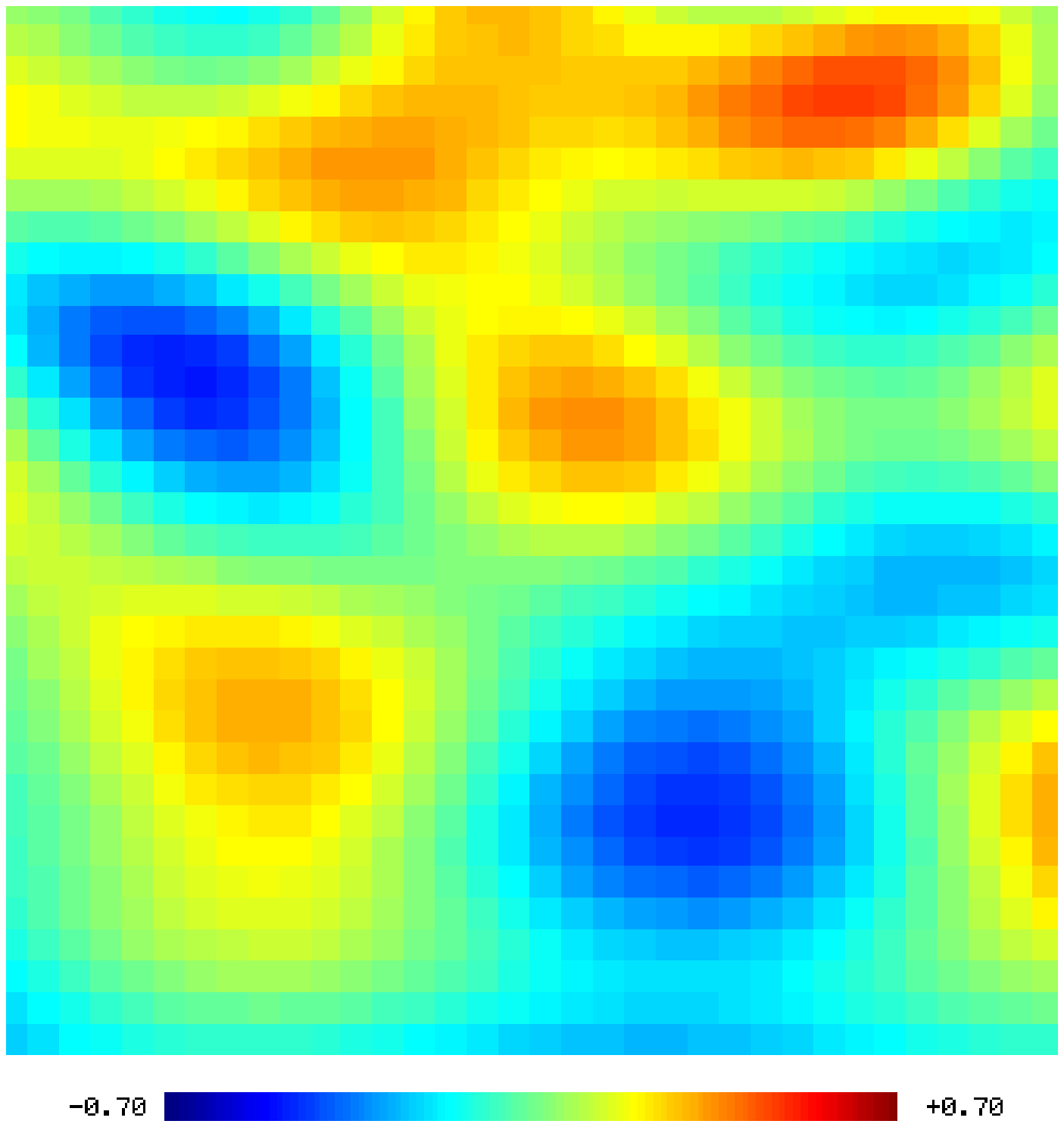}}
    \caption{Mosaic correlation map of the ILC signal with the Faraday rotation depth in the CS region. The window is $2^\circ \times 2^\circ$. (left panel). Right panel. Smoothed map of the correlation between the ILC signal and the Faraday rotation depth.}
    \label{fig_cs_corr}
  \end{center}
\end{figure}

\begin{figure}
  \begin{center}
    \centerline{\includegraphics[width=0.35\linewidth, angle=-90]{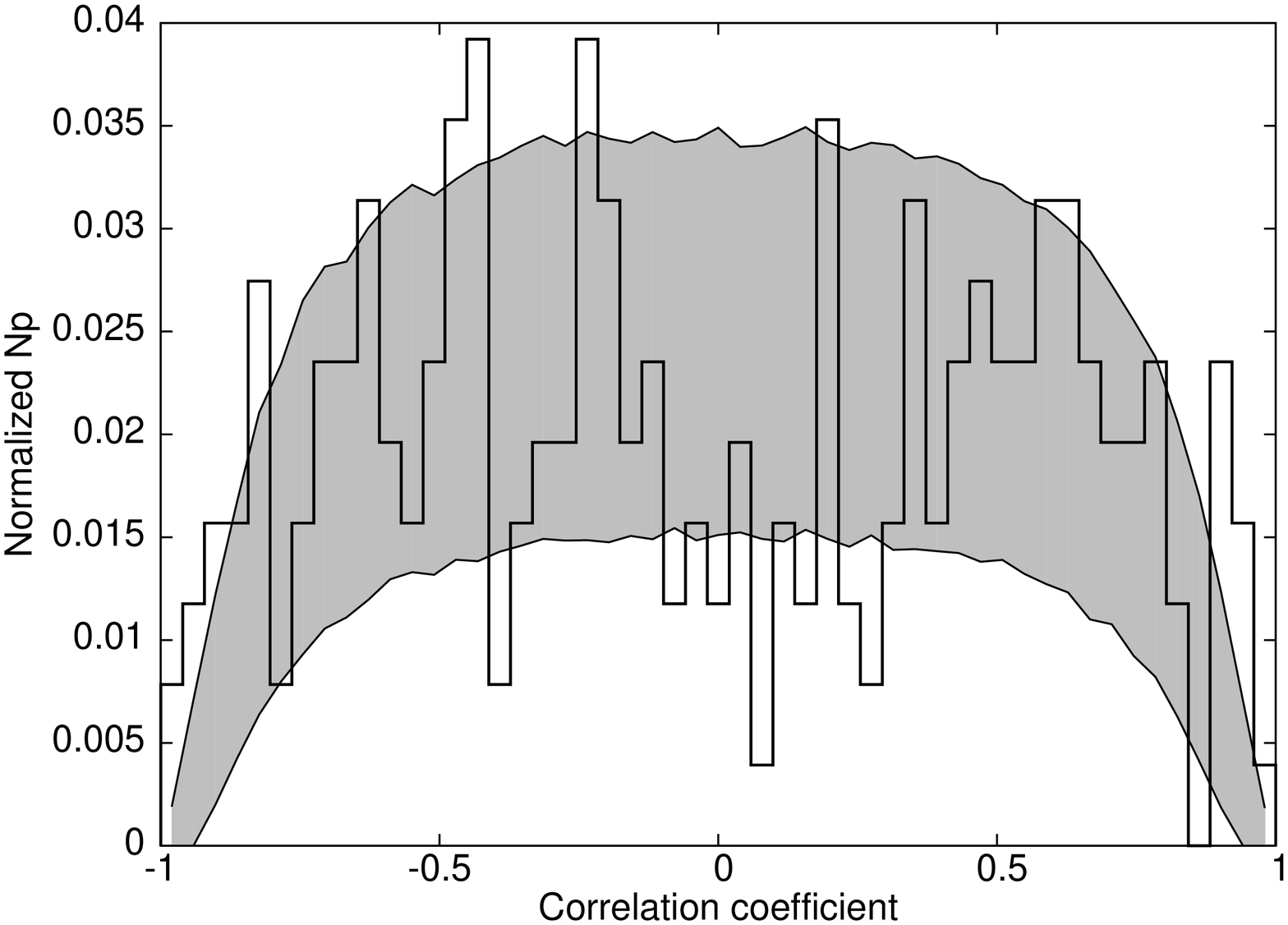}
                \includegraphics[width=0.35\linewidth, angle=-90]{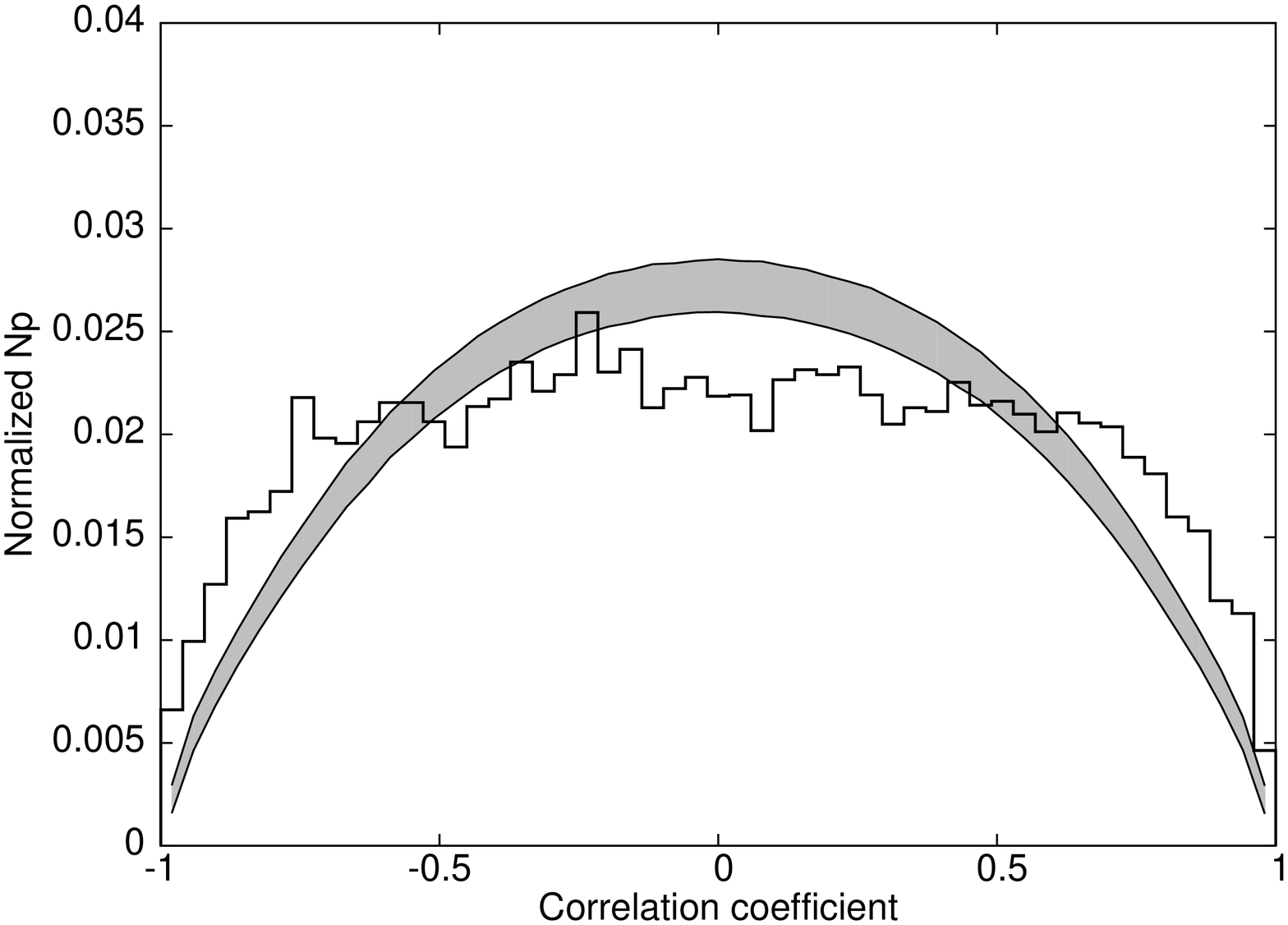}}
    \caption{Histogram of the correlation coefficients in the CS region (the left panel), and for the entire CMB sky (the right panel). The correlation window is $2^\circ \times 2^\circ$. The black solid line is the normalized number of pixels for the ILC map. The gray area demonstrate the $\pm1\sigma$-scatter produced for 1000 simulations.}
    \label{fig_hist_cs}
  \end{center}
\end{figure}

We now turn to the mosaic correlation method \citep{Mosaic}, to study the correlation of the Faraday depth signal in the CS zone with the WMAP ILC data.\\
For any two maps with the same pixelization and angular resolution, we define the mosaic cross-correlation $K(l,b)$ as:
\begin{eqnarray}
K(l,b) &=& \frac{1}{\sigma_x\sigma_y}\sum_{i\in\Omega(l,b)}\left(x_i-<x_i>)(y_i-<y_i>\right),\nonumber\\
\sigma^2_x &=& \sum_{i \in \Omega(l,b)}(x_i-<x_i>)^2,\nonumber\\
\sigma^2_y &=& \sum_{i \in \Omega(l,b)}(y_i-<y_i>)^2
\label{eq5}
\end{eqnarray}
where $l,b$ denote the coordinates of the zone $\Omega(l,b)=S^\circ \times S^\circ$, with pixel numbers $i \in \Omega$ and $<..>$ denote the average over all pixels $i$ in the zone $\Omega(l,b)$.\\
Using $K(l,b)$ from Eq.\ref{eq5}, we now produce a mosaic correlation of the ILC map and the Faraday depth measure map. In Fig. \ref{fig_cs_corr} we present a cut of $30^\circ \times 30^\circ$, around the cold spot area, using a correlation window of $\Omega(l,b)=2^\circ\times2^\circ$.\\
Here, the blue spots corresponds to anticorrelational zones of the ILC and Faraday depth signal, and the red spots indicate  zones with positive correlations. We calculated statistics of these correlational coefficients and compared it with 1000 simulations of the Faraday rotation and CMB model correlational data. The histogram on Fig. \ref{fig_hist_cs} show the distribution of the coefficients in the CS region (left), and for the total sky (right). Notice the two-peaks distribution, different from the expected for an uncorrelated signal. Such a distribution is connected with the close positions of maxima and minima in the ILC and Faraday depth maps.

\begin{figure}
  \begin{center}
    \centerline{\includegraphics[width=0.80\linewidth]{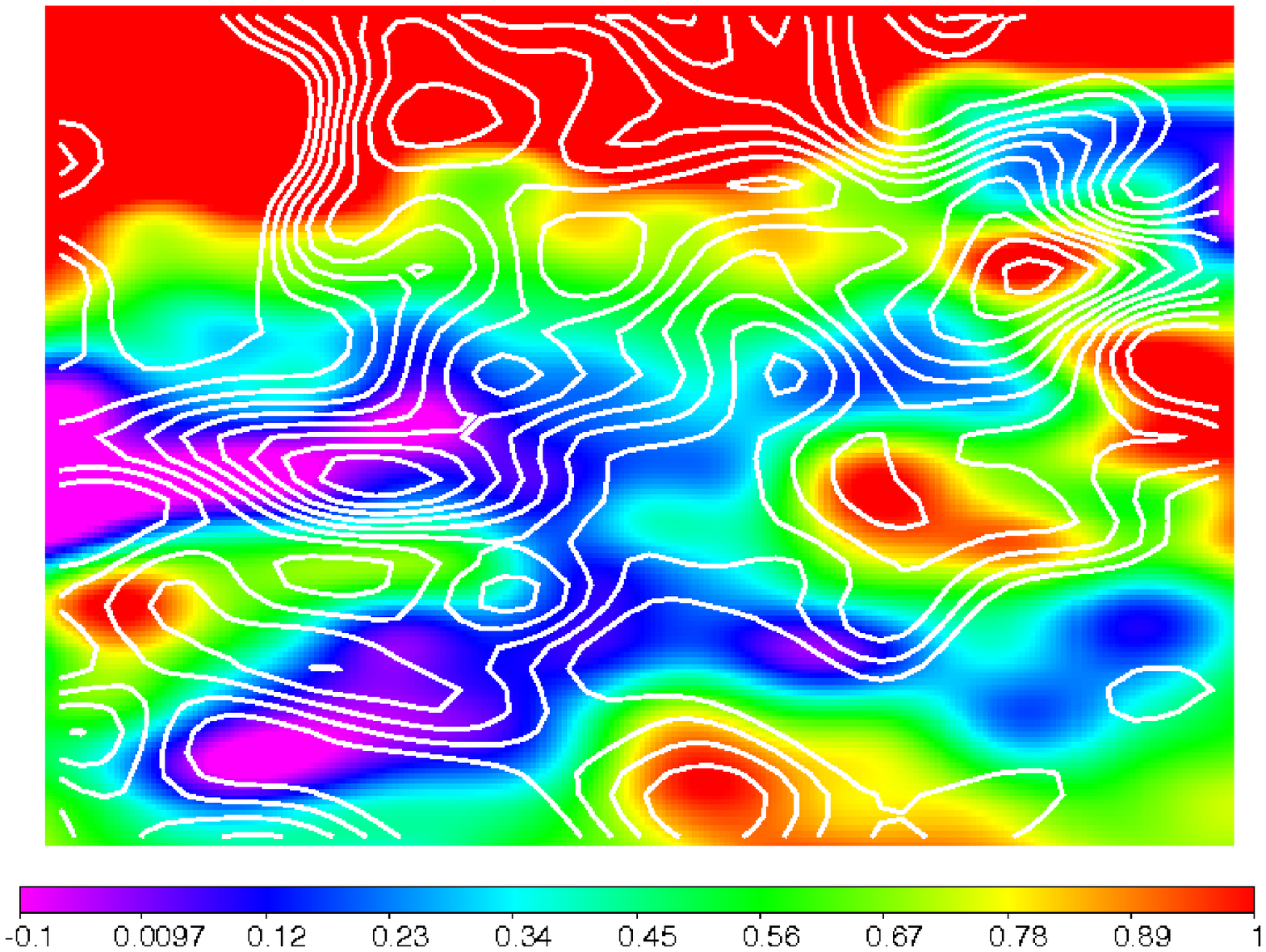}}
    \centerline{\includegraphics[width=0.80\linewidth]{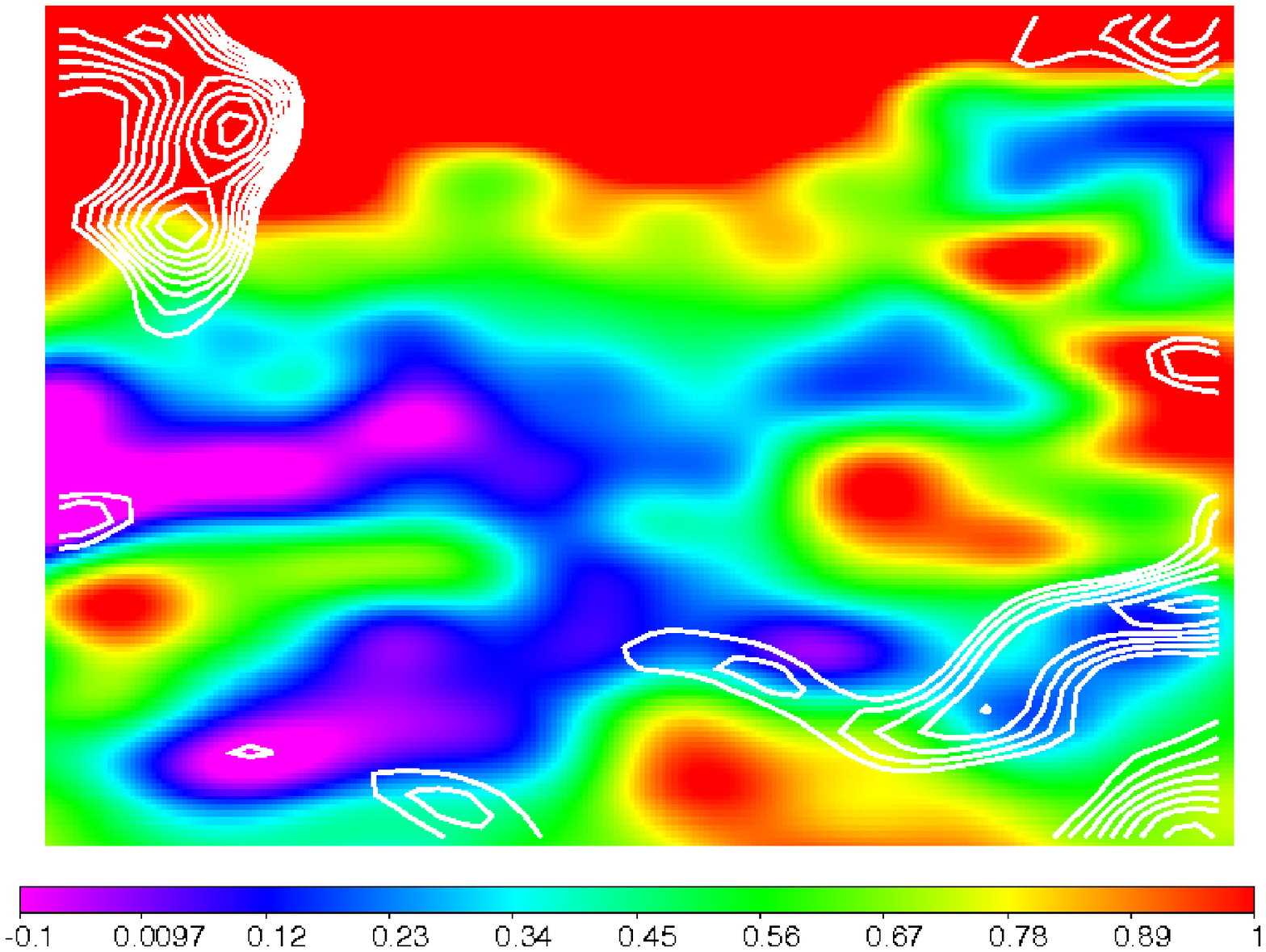}}
    \caption{Similar to Fig. \ref{fig_FC1}, but for the Cold Spot region ($l=209^{\circ}$, $b=-57^{\circ}$).
}
    \label{fig_cs_iso_neg}
  \end{center}
\end{figure}

Finally, similar to the procedure above, we test the area around the CS for overlap between the extrema of the Faraday depth and of the CMB temperature. The results are presented in Fig. \ref{fig_cs_iso_neg}. It is clear to see, in the top panel, that the concentration of isolines and the areas of high and low Faraday depth overlap well. The high temperature isolines in the right panel does not fit as good, although there is some overlap in the lower right area of the sky patch. Nonetheless, these results are even stronger, than those from Figs. \ref{fig_FC1} through \ref{fig_FC4}, and underline the peculiarity of the high correlations in the CS region.

\section{Correlation at the Galactic radio loops}
\begin{figure}
  \begin{center}
    \centerline{\includegraphics[width=1.0\linewidth]{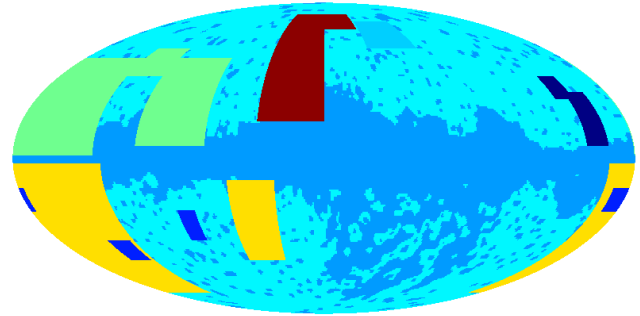}}
    \caption{The 6 galactic radio loops, with the kq75p1 mask as background. Dark red: Loop 1 A \& B. Yellow: Loop II A \& B. Light green: Loop III A, B \& C. Cyan: Loop IV A. Blue: Loop V A, B \& C. Dark blue: Loop VI A \& B.}
    \label{fig:loops} 
  \end{center}
\end{figure}

\begin{figure}
  \begin{center}
    \centerline{\includegraphics[width=1.0\linewidth]{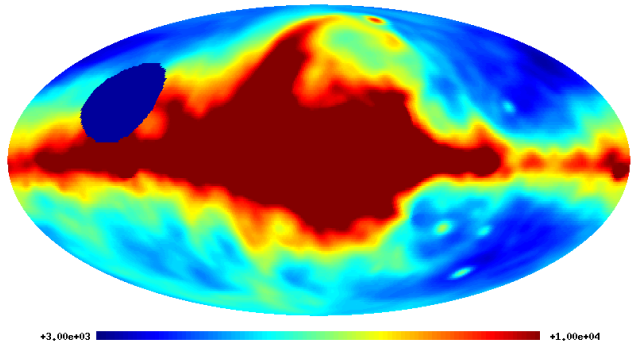}}
    \caption{The sky at 45 MHz, with the north galactic spur clearly visible. Color scale is chosen to emphasize the spur structure. The blue circular area to the left of the map is not covered in the dataset.}
    \label{fig:spur} 
  \end{center}
\end{figure}

We now turn to areas on the map, at high and low galactic latitude, covered by the galactic radio loops. 
The shock front of the super nova remnants (SNR), outlined by the galactic loops, affects the observed galactic magnetic field. The magnetic fields at the shock front may be the origin of synchrotron radiation, possibly affecting the CMB. To investigate, whether or not the area of the SNRs affect the CMB, we cross correlate the Faraday $\phi$-map and the ILC7 map, at the loop locations. The areas in the sky occupied by the loops are shown in fig. \ref{fig:loops} with the kq75p1 mask as background, and in table \ref{table:loops}, with the limiting longitudes and latitudes taken from \citep{Borka}. The area covered by loop I, is overlapping the well known northern galactic spur, which is very clearly visible at 45 MHz in Fig. \ref{fig:spur}. The data for the 45 MHz map is taken from \citep{Guzman}.

\begin{table}
\begin{center}
\caption{The galactic longitude ($l$) and latitude ($b$) in degrees, for the loops I-VI.}
\begin{tabular}{| c | c  l  l |}
\hline
\multicolumn{2}{| c }{Loop} & $l$ ($^{\circ}$) & $b$ ($^{\circ}$)\\
\hline
\multirow{2}{*}{Loop I}   & A & [40,0] & [18,78] \\
                          & B & [360,327] & [67,78] \\
\hline
\multirow{2}{*}{Loop II}  & A & [57,30] & [-50,-10] \\
                          & B & [195,130] & [-70,-2] \\
\hline
\multirow{3}{*}{Loop III} & A & [180,135] & [2,50] \\
                          & B & [135,110] & [40,55] \\
                          & C & [110,70] & [6,50] \\
\hline
\multirow{1}{*}{Loop IV}  & A & [325,285] & [55,72] \\
\hline
\multirow{3}{*}{Loop V}   & A & [189,178] & [-25,-13] \\
                          & B & [147,133] & [-50,-39] \\
                          & C & [90,80] & [-39,-24] \\
\hline
\multirow{2}{*}{Loop VI}  & A & [215,205] & [29,40] \\
                          & B & [207,196] & [6,32]\\
\hline
\end{tabular}
\end{center}
\label{table:loops}
\end{table}

\begin{figure}
  \begin{center}
    \centerline{\includegraphics[width=1.0\linewidth]{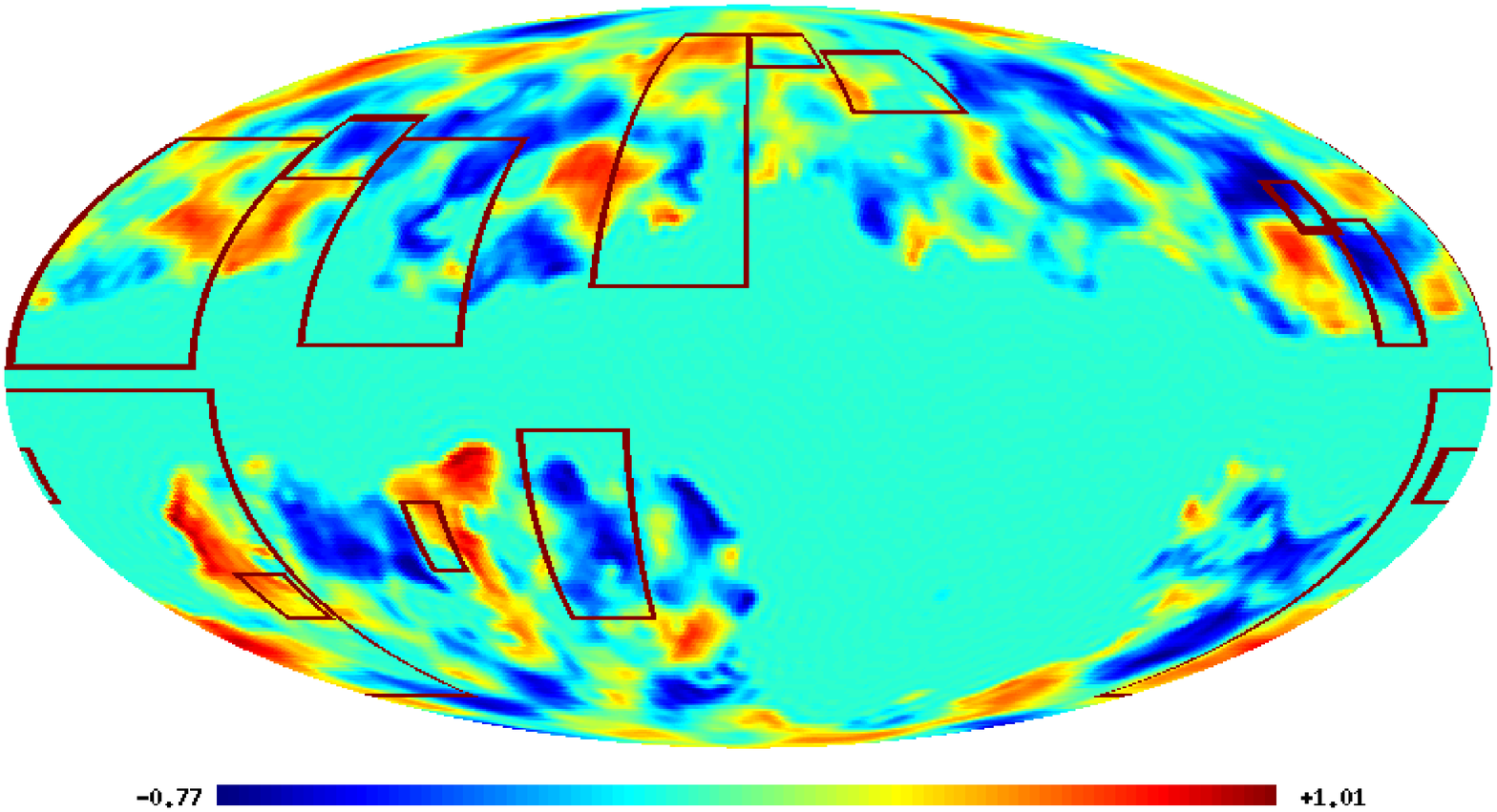}}
    \centerline{\includegraphics[width=1.0\linewidth]{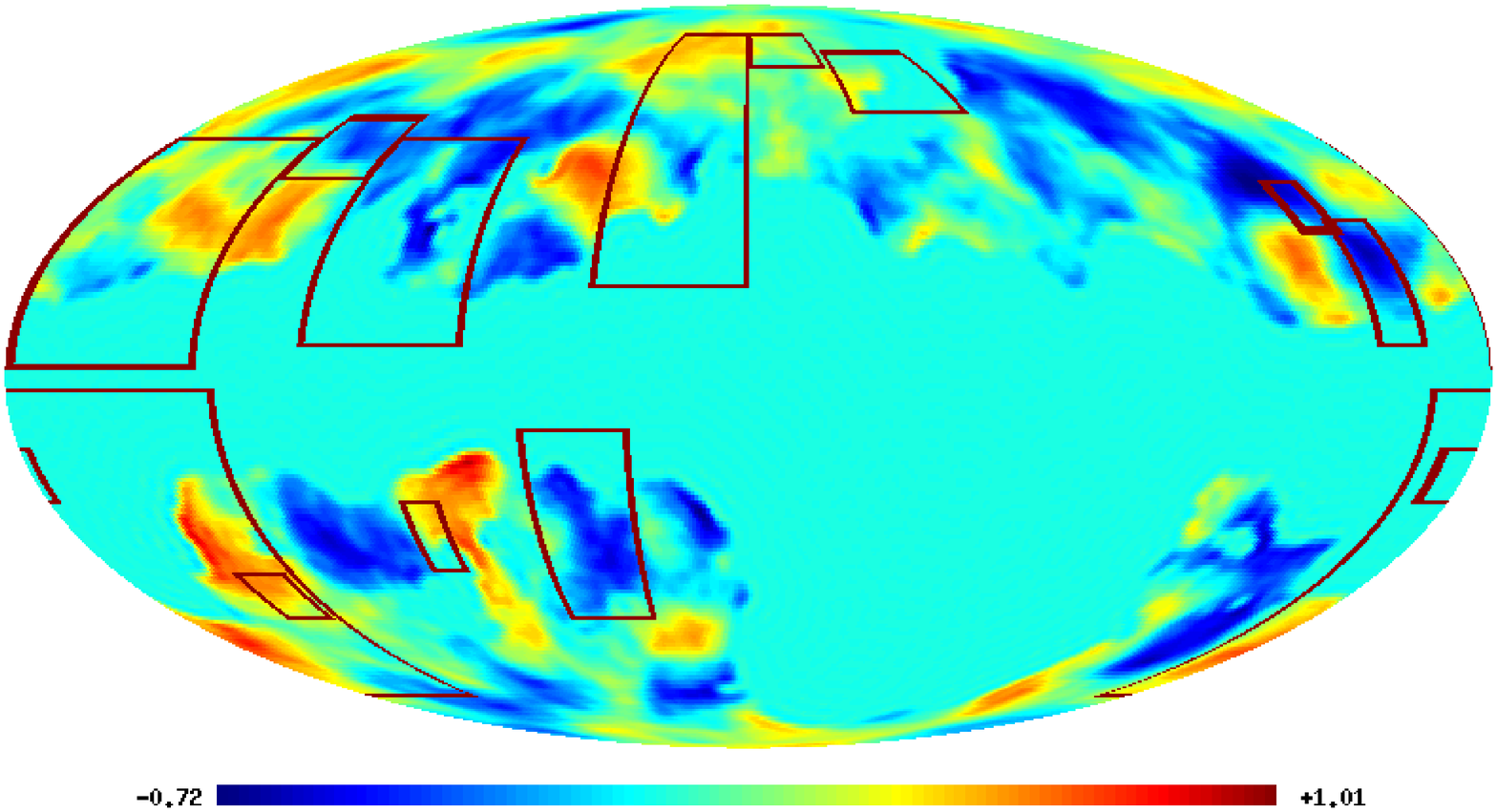}}
    \caption{The contours of the 6 galactic radio loops, overlaid with the $r_{s}$ map for $R=5^{\circ}$ (top) and $R=7^{\circ}$ (bottom).}
    \label{fig:contours}
  \end{center}
\end{figure}

In Fig. \ref{fig:contours} we overlay the areas of the radio loops with the $r_{s}$ maps from Figs. \ref{ilc7-5deg} and \ref{ilc7-7deg}. It is interesting, that several of the areas of high and low correlation are located inside the radio loops. Most interesting are loop II A (in the middle lower part) and loop VI A \& B (right upper part), which coinside very well with areas of negative correlation. Of areas with positive correlation, loop V B and somewhat C (the second and first small loop, left of II A respectively) stand out. It is clear, that there is no preference for either negative or positive correlation in the loops, as will also be shown below. However the loops do, for the most part, contain areas of strong correlation.

We now investigate the correlation between the Faraday $\phi$-map, and the CMB ILC7 map, and the V-band map in particular, in the areas of the galactic radio loops. Again we use the method of local cross-correlation, introduced earlier, only for square areas on the sky this time. In order to minimize the contribution to the correlation, caused by noise in the two maps, we smooth them, before calculating the cross correlation. We select smoothing values of $1^{\circ}$, $5^{\circ}$ and $10^{\circ}$, for both the $\phi$ map, as well as the ILC7 and V-band map. We then apply the kq75p1 mask to the $\phi$-map and the ILC7 and V-band map, which we have scaled to the same resolution as the mask. We only consider pixels, not covered by the mask, for the calculation of the cross correlations. Note, that one of the loop V areas (A) are almost completely covered by the mask, resulting in extreme values of the correlations.\\
Finally, we compare the cross correlation value for each loop area, with the cross correlation value of 100 simulated CMB maps, and count the amount of simulations having a greater value of $r_s$. The entire procedure described above, was conducted using the GLESP program \citep{Glesp}, with a resolution of $n_{x}=201$, meaning that the map had 201 pixels in the $\phi$-direction, and 402 in the $\theta$-direction.\\
The results from the cross correlations can be seen in tables \ref{table:results} and \ref{table:results_v}, for smoothings of $1^{\circ}$, $5^{\circ}$ and $10^{\circ}$. Also included in the tables, is the number, $S$, of simulations with a higher value of the cross-correlations.

\begin{table}
\begin{center}
\caption{The cross correlation ($r_s$) for the galactic radio loops, and the amount of simulations with a higher cross correlation ($S$), for three values of smoothing. This is compared to 100 simulations.}
\begin{tabular}{| c  c | l  l | l  l | l  l |}
\hline
\multicolumn{2}{| c }{ Loop } & \multicolumn{2}{| c }{ Smooth $1^{\circ}$ } & \multicolumn{2}{| c }{ Smooth $5^{\circ}$ } & \multicolumn{2}{| c |}{ Smooth $10^{\circ}$ } \\
\multicolumn{2}{| c }{ } & \multicolumn{1}{| c }{$r_s$} & \multicolumn{1}{ c }{$S$} & \multicolumn{1}{| c }{$r_s$} & \multicolumn{1}{ c }{$S$} & \multicolumn{1}{| c }{$r_s$} & \multicolumn{1}{ c |}{$S$}\\
\hline
\multirow{2}{*}{Loop I}   & A & 0.534 & 15 & 0.737 & 8 & 0.807 & 5 \\
                          & B & 0.760 & 5 & 0.888 & 10 & 0.961 & 17 \\
\hline
\multirow{2}{*}{Loop II}  & A & -0.323 & 93 & -0.409 & 89 & -0.456 & 85 \\
                          & B & 0.335 & 9 & 0.377 & 10 & 0.401 & 14 \\
\hline
\multirow{3}{*}{Loop III} & A & 0.283 & 7 & 0.445 & 7 & 0.573 & 6 \\
                          & B & 0.108 & 37 & 0.553 & 19 & 0.903 & 5 \\
                          & C & 0.127 & 35 & 0.328 & 29 & 0.615 & 21 \\
\hline
\multirow{1}{*}{Loop IV}  & A & 0.330 & 33 & 0.392 & 43 & 0.447 & 50 \\
\hline
\multirow{3}{*}{Loop V}   & A & 0.999 & 46 & -0.675 & 54 & -0.996 & 54 \\
                          & B & 0.068 & 48 & 0.600 & 28 & 0.930 & 1 \\
                          & C & 0.208 & 51 & 0.259 & 51 & 0.165 & 54 \\
\hline
\multirow{2}{*}{Loop VI}  & A & -0.0311 & 49 & 0.119 & 49 & 0.649 & 40 \\
                          & B & 0.049 & 50 & 0.233 & 46 & 0.692 & 27 \\
\hline
\end{tabular}
\end{center}
\label{table:results}
\end{table}

\begin{table}
\begin{center}
\caption{Similar to table \ref{table:results}, but for the V-band only.}
\begin{tabular}{| c  c | l  l | l  l | l  l |}
\hline
\multicolumn{2}{| c }{ Loop } & \multicolumn{2}{| c }{ Smooth $1^{\circ}$ } & \multicolumn{2}{| c }{ Smooth $5^{\circ}$ } & \multicolumn{2}{| c |}{ Smooth $10^{\circ}$ } \\
\multicolumn{2}{| c }{ } & \multicolumn{1}{| c }{$r_s$} & \multicolumn{1}{ c }{$S$} & \multicolumn{1}{| c }{$r_s$} & \multicolumn{1}{ c }{$S$} & \multicolumn{1}{| c }{$r_s$} & \multicolumn{1}{ c |}{$S$}\\
\hline
\multirow{2}{*}{Loop I}   & A & 0.532 & 15 & 0.754 & 8 & 0.820 & 3 \\
                          & B & 0.731 & 5 & 0.887 & 10 & 0.963 & 17 \\
\hline
\multirow{2}{*}{Loop II}  & A & -0.308 & 93 & -0.402 & 88 & -0.459 & 85 \\
                          & B & 0.327 & 9 & 0.367 & 12 & 0.384 & 15 \\
\hline
\multirow{3}{*}{Loop III} & A & 0.276 & 8 & 0.447 & 7 & 0.574 & 6 \\
                          & B & 0.110 & 37 & 0.580 & 19 & 0.920 & 1 \\
                          & C & 0.117 & 35 & 0.347 & 29 & 0.601 & 22 \\
\hline
\multirow{1}{*}{Loop IV}  & A & 0.345 & 31 & 0.482 & 36 & 0.578 & 46 \\
\hline
\multirow{3}{*}{Loop V}   & A & 0.999 & 44 & -0.027 & 53 & -0.990 & 53 \\
                          & B & 0.162 & 43 & 0.669 & 20 & 0.930 & 1 \\
                          & C & 0.219 & 51 & 0.239 & 52 & 0.0798 & 54 \\
\hline
\multirow{2}{*}{Loop VI}  & A & -0.053 & 49 & 0.092 & 49 & 0.655 & 40 \\
                          & B & 0.026 & 51 & 0.231 & 46 & 0.746 & 23 \\
\hline
\end{tabular}
\end{center}
\label{table:results_v}
\end{table}

\begin{figure}
  \begin{center}
    \centerline{\includegraphics[width=1.0\linewidth]{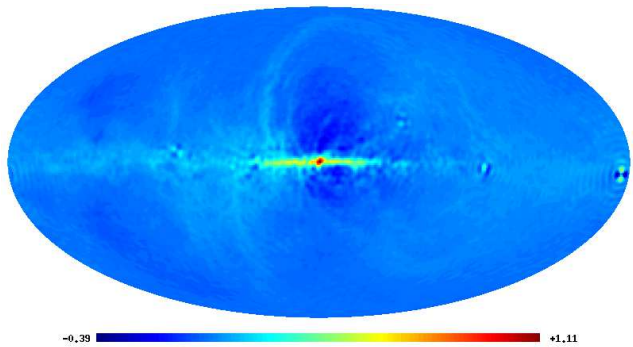}}
    \caption{The map of E-modes in the K-band. Notice the spur-like structure, extending from the left of the galactic centre and extending northwards.}
    \label{fig_E_mode}
  \end{center}
\end{figure}

From the table of results, it is immediately obvious, that the smoothing have a big effect on certain loops. Because of the contribution from areas outside the loops during the smoothing process, the small area loops are naturally most affected (see for instance loop III B). Further, there is not much difference between the ILC7, and the V-band, meaning, that the ILC method does not cause any of the correlation effects. We also see, that for almost all loops, we have an increase in the $r_s$, when we increase the smoothing degree, as the contribution from noise diminishes. However, only some of the areas show a systematic increase in significance compared to the random simulations. Of particular interest are the areas covered by loop I A, and loop III A, as these areas cover a relatively large area of the sky, and consistently exhibit a strong correlation compared to simulations. Further, as Fig. \ref{fig:spur} shows, the boundary of Loop I could be extended left, to more accurately overlap the north galactic spur. From Fig.\ref{fig:contours} we see, that this extension would incorporate two bright red spots into the correlation coefficient, and thus increase the level of cross-correlations above the threshold mentioned above.\\
Although the results are not very strong, this nonetheless shows, that the magnetic field in the powerful radio loops may affect the CMB photons.\\
Further, we have also investigated the correlations in the loop for the CMB polarization. We have used the E-mode polarization from the K-band for the test, and compared to simulations created via Healpix \citep{healpix}. Apart from the different maps, the procedure for the correlations in the loops is exactly the same, as described above.\\
In Fig. \ref{fig_E_mode} we show the E-modes from the K-band. Notice the spur-like structure, extending from the left of the galactic centre and extending northwards. This have a remarkable resemblance to a similar structure, visible in roughly the same area of the map in Fig. \ref{s-map}, though it is not clearly visible in the $\phi$-map (Fig \ref{phi-map}). And additionally, the area coincide reasonably well with the area of Loop I, as seen in Fig. \ref{fig:loops}, making it particularly interesting.\\

The results of the correlation test for the polarization are presented in table \ref{table:results_E}. In overview, many results seem significant at $1^{\circ}$ smoothing, but becomes non-significant at greater levels of smoothings. Also interesting is the seemingly high levels of correlation, where several areas are distinctly anti-correlated.\\
The area mentioned above, at Loop I (A), like several other loop areas, shows a significant anti-correlation at $1^{\circ}$ smoothing, but at greater smoothings the correlation value quickly drops to insignificant levels. This could be explained by the fact, that the spur-like structure in Fig. \ref{fig_E_mode} seems to be relatively thin. At greater levels of smoothing, the feature becomes indistinct, and thus the significance drops.\\
A more detailed analysis of these correlations will be the aim for future work. Presumably, when higher quality polarization maps, as well as better data for the Faraday depth will be available, the results may improve. 

\begin{table}
\begin{center}
\caption{Similar to table \ref{table:results}, but for the K-band, E-mode polarization, compared with simulated polarization data.}
\begin{tabular}{| c  c | l  l | l  l | l  l |}
\hline
\multicolumn{2}{| c }{ Loop } & \multicolumn{2}{| c }{ Smooth $1^{\circ}$ } & \multicolumn{2}{| c }{ Smooth $5^{\circ}$ } & \multicolumn{2}{| c |}{ Smooth $10^{\circ}$ } \\
\multicolumn{2}{| c }{ } & \multicolumn{1}{| c }{$r_s$} & \multicolumn{1}{ c }{$S$} & \multicolumn{1}{| c }{$r_s$} & \multicolumn{1}{ c }{$S$} & \multicolumn{1}{| c }{$r_s$} & \multicolumn{1}{ c |}{$S$}\\
\hline
\multirow{2}{*}{Loop I}   & A & -0.712 & 99 & -0.732 & 83 & -0.726 & 78 \\
                          & B & -0.622 & 90 & -0.763 & 77 & -0.831 & 73 \\
\hline
\multirow{2}{*}{Loop II}  & A & 0.099 & 36 & 0.120 & 43 & 0.154 & 44 \\
                          & B & 0.143 & 28 & 0.150 & 40 & 0.151 & 43 \\
\hline
\multirow{3}{*}{Loop III} & A & -0.036 & 56 & -0.031 & 49 & -0.052 & 49 \\
                          & B & -0.584 & 98 & -0.649 & 78 & -0.708 & 76 \\
                          & C & -0.603 & 100 & -0.687 & 85 & -0.770 & 84 \\
\hline
\multirow{1}{*}{Loop IV}  & A & -0.629 & 98 & -0.759 & 84 & -0.857 & 78 \\
\hline
\multirow{3}{*}{Loop V}   & A & -0.076 & 58 & -0.068 & 58 & -0.066 & 56 \\
                          & B & 0.526 & 3 & 0.615 & 22 & 0.677 & 27 \\
                          & C & 0.529 & 14 & 0.545 & 37 & 0.544 & 38 \\
\hline
\multirow{2}{*}{Loop VI}  & A & -0.642 & 89 & -0.648 & 64 & -0.643 & 63 \\
                          & B & -0.590 & 91 & -0.622 & 73 & -0.643 & 68 \\
\hline
\end{tabular}
\end{center}
\label{table:results_E}
\end{table}

\section{Conclusion}
In this paper, we have investigated possible foreground contaminations of the CMB, at high and low galactic latitudes. We have cross correlated CMB maps, with the Faraday Rotation map, to identify areas of high or low correlation. Specifically, in our first test, we selected circular areas, of various radii for our comparison test. The idea is, that the Faraday Rotation is affected by thermal electrons, as well as the strength of the galactic magnetic field. A large value for the Faraday Rotation, at a given area, can imply that we have a high column density of thermal electrons, and thus create a cold area in the CMB. Our cross correlation test for the ILC map showed that there was a considerable anti-correlation between the Faraday rotation map and the ILC map, at the Cold Spot (CS), as well as in a number of other areas, all well outside the galactic plane. The correlations in the CS area were compared to correlations over the rest of the sky, where the results showed the CS area as highly anticorrelated. We further tested the area around the CS with the mosaic correlation method, with a $2^{\circ} \times 2^{\circ}$ window. \\
In a histogram of the the number of pixels contra correlation coefficients for the mosaic correlation, we found two significant peaks compared to random simulations, at negative correlation values, in good agreement with the results for the cross correlations.\\
Additionally, we compared the Faraday depth and the CMB temperature isolines, in an area around the CS and at 4 other locations of high correlation value. We found in general, that many of the CMB temperature extrema, and Faraday depth extrema overlapped, substantiating the theory that the magnetic field affects the CMB temperature. Note that this were all for areas well outside the galactic plane.\\
Lastly, we compared our maps of correlation with simulated maps, and found that the standard deviation for the mean cross correlation in the ILC case, deviated from random simulations at the $99.9\%$ level.\\
For our second test, we repeated the cross correlations, but this time at very specific areas of the map. We focused on the galactic radio loops, which is the remnants of supernova explosions. The shock fronts of the supernovas, outlined by the galactic radio loops, contains strong magnetic fields, possibly giving rise to synchrotron radiation. Thus we cross correlated the Faraday Rotation maps, in the areas of the galactic radio loops, and compared these with random simulations. During the correlations, we smoothed both maps to various degrees, to avoid contamination from the noise. The results for these tests were not as strong, as for the unspecified area test above, but nonetheless interesting. For the ILC case, we found the most powerful results for the heavy smoothing of 10 $^{\circ}$, where loop I(A) and loop III(B) both had a significance at the $95\%$ level. The area occupied by loop III(A) was also interesting, having a relatively high significance, irrespective of the smoothing. We repeated the test for the V-band, but found similar results, showing that the process of creating the ILC map, is not the cause of the correlations.
Finally, we also investigated the correlation between the Faraday depth map and the K-band E-mode polarization, in the loop areas. The results showed, that we had significant anticorrelations for the case of $1^{\circ}$ smoothing, but insignificant correlations for greater levels of smoothing.\\
Comon to all the tests are, that at current, the Faraday maps used are not of the highest quality, and thus the results for the comparisons may improve in the future.\\
In conclusion, we find that foreground contamination at high galactic lattitudes, should be taken very seriously, and if not properly filtered, may cause anomalies in the CMB.

\section{Acknowledgments}
We acknowledge the use of the GLESP package \citep{Glesp} to produce CMB maps, and the use of Healpix \citep{healpix}. O.V.V. thanks Dmitry Zimin's nonprofit Dynasty Foundation for the support. This work is supported in part by Danmarks Grundforskningsfond, which allowed the establishment of the Danish Discovery Center. This work is supported by FNU grant 272-06-0417, 272-07-0528 and 21-04-0355.


\begin{thebibliography}{}

\bibitem[\protect\citeauthoryear{Bennett et al.}{2003}]{wmapresults}
C. L. Bennett, et al.
\newblock{First-Year Wilkinson Microwave Anisotropy Probe (WMAP) Observations: Preliminary Maps and Basic Results.}
\newblock{\em The Astrophysical Journal Supplement}, 148, 1, 2003

\bibitem[\protect\citeauthoryear{Bennett et al.}{2011}]{wmap7ylowl}
C. L. Bennett, et al.
\newblock{Seven-year Wilkinson Microwave Anisotropy Probe (WMAP) Observations: Are There Cosmic Microwave Background Anomalies?}
\newblock{\em The Astrophysical Journal Supplement}, 192, 17, 2011

\bibitem[\protect\citeauthoryear{Borka}{2007}]{Borka}
V. Borka.
\newblock{Spectral indices of Galactic radio loops between 1420, 820 and 408 MHz.}
\newblock{\em MNRAS}, 376, 634, 2007

\bibitem[\protect\citeauthoryear{Cayon et al.}{2005}]{cjt}
L. Cayon, J. Jin and A. Treaster.
\newblock{Higher Criticism Statistic: Detecting and Identifying Non-Gaussianity in the WMAP First Year Data.}
\newblock{\em MNRAS}, 362, 826, 2005

\bibitem[\protect\citeauthoryear{Condon et al. }{1998}]{nvss}
J. J. Condon, et al.
\newblock{The NRAO VLA Sky Survey.}
\newblock{\em Astron. J.}, 115, 1693, 1998, 

\bibitem[\protect\citeauthoryear{Cruz et al.}{2005}]{cruz2005}
M. Cruz, E. Martinez-Gonzales, P. Vielva and L. Cayon.
\newblock{Detection of a non-Gaussian Spot in WMAP.}
\newblock{\em MNRAS},356, 29, 2005

\bibitem[\protect\citeauthoryear{Cruz et al.}{2006}]{cruz2006}
M. Cruz, M. Tucci, E. Martinez-Gonzales and P. Vielva.
\newblock{The non-Gaussian cold spot in Wilkinson Microwave Anisotropy Probe: significance, morphology and foreground contribution.}
\newblock{\em MNRAS}, 369, 57, 2006

\bibitem[\protect\citeauthoryear{Cruz et al.}{2007}]{cruz2007}
M. Cruz, L. Cayon, E. Martinez-Gonzales, P. Vielva and J. Jin.
\newblock{The non-Gaussian Cold Spot in the 3-year WMAP data.}
\newblock{\em The Astrophysical Journal}, 655, 11, 2007

\bibitem[\protect\citeauthoryear{Cruz et al.}{2008}]{Turok}
M. Cruz, et al.
\newblock{The CMB cold spot: texture, cluster or void?}
\newblock{\em MNRAS}, 390, 913, 2008

\bibitem[\protect\citeauthoryear{Das \& Spergel}{2009}]{Spergel}
S. Das and D. N. Spergel.
\newblock{CMB lensing and the WMAP cold spot.}
\newblock{\em Phys. Rev. D}, 79, 043007, 2009

\bibitem[\protect\citeauthoryear{Dineen \& Coles}{2004}]{dineen2004}
P. Dineen, and P. Coles.
\newblock{Faraday rotation as a diagnostic of Galactic foreground contamination of cosmic microwave background maps.}
\newblock{\em MNRAS}, 347, 52, 2004

\bibitem[\protect\citeauthoryear{Guzman}{2011}]{Guzman}
A. E. Guzman, J. May, H. Alvarez and K. Maeda.
\newblock{All-sky Galactic radiation at 45 MHz and spectral index between 45 and 408 MHz.}
\newblock{\em A \& A}, 525, A138, 2011

\bibitem[\protect\citeauthoryear{Doroshkevich et al.}{2005}]{Glesp}
A. G. Doroshkevich, P. D. Naselsky, O. V. Verkhodanov, D. I. Novikov, V. I. Turchaninov, I. D. Novikov, P. R. Christensen and L. -Y. Chiang.
\newblock{Gauss-Legendre Sky Pixelization (GLESP) for CMB maps.}
\newblock{\em International Journal of Modern Physics D}, 14, 275-290, 2005, http://www.glesp.nbi.dk/

\bibitem[\protect\citeauthoryear{Gold et al.}{2011}]{gold2011}
B. Gold, et al.
\newblock{Seven-year Wilkinson Microwave Anisotropy Probe (WMAP) Observations: Galactic Foreground Emission.}
\newblock{\em The Astrophysical Journal Supplement}, 192, 15, 2011

\bibitem[\protect\citeauthoryear{Gorski et al.}{2005}]{healpix}
K. M. Gorski et al.
\newblock{HEALPix: A Framework for High-Resolution Discretization and Fast Analysis of Data Distributed on the Sphere.}
\newblock{\em The Astrophysical Journal}, 622, 759, 2005

\bibitem[\protect\citeauthoryear{Hinshaw et al.}{2007}]{wmap3ytem}
G. Hinshaw, et al.
\newblock{Three-Year Wilkinson Microwave Anisotropy Probe (WMAP) Observations: Temperature Analysis.}
\newblock{\em The Astrophysical Journal Supplement}, 170, 288, 2007

\bibitem[\protect\citeauthoryear{Hinshaw et al.}{2009}]{wmap5ytem}
G. Hinshaw, et al.
\newblock{Five-Year Wilkinson Microwave Anisotropy Probe (WMAP) Observations: Data Processing, Sky Maps, and Basic Results.}
\newblock{\em The Astrophysical Journal Supplement}, 180, 225, 2009 

\bibitem[\protect\citeauthoryear{Jarosik, et al.}{2011}]{wmap7ytem}
N. Jarosik, et al.
\newblock{Seven-Year Wilkinson Microwave Anisotropy Probe (WMAP) Observations: Sky Maps, Systematic Errors, and Basic Results.}
\newblock{\em The Astrophysical Journal Supplement}, 192, 14, 2011

\bibitem[\protect\citeauthoryear{Kobulnicky \& Johnson}{1999}]{kobulnicky}
H. A. Kobulnicky and K. E. Johnson.
\newblock{Signatures of the Youngest Starbursts: Optically Thick Thermal Bremsstrahlung Radio Sources in Henize 2-10.}
\newblock{\em The Astrophysical Journal}, 527, 154, 1999

\bibitem[\protect\citeauthoryear{Komatsu et al.}{2011}]{komatsu2011}
E. Komatsu et al. 
\newblock{Seven-year Wilkinson Microwave Anisotropy Probe (WMAP) Observations: Cosmological Interpretations.}
\newblock{\em The Astrophysical Journal Supplement}, 192, 18, 2011

\bibitem[\protect\citeauthoryear{Large et al.}{1962}]{Large1}
M. I. Large, M. J. S. Quigley and C. G. T. Haslam.
\newblock{A new feature of the radio sky.}
\newblock{\em MNRAS}, 124, 405, 1962

\bibitem[\protect\citeauthoryear{Large et. al.}{1966}]{Large2}
M. I. Large, M. J. S. Quigley and C. G. T. Haslam.
\newblock{A radio study of the north polar spur. II, A survey at low declinations.}
\newblock{\em MNRAS}, 131, 335, 1966

\bibitem[\protect\citeauthoryear{McKee \& Ostriker}{1977}]{mckee}
C. F. McKee and J. P Ostriker.
\newblock{A theory of the interstellar medium - Three components regulated by supernova explosions in an inhomogeneous substrate.}
\newblock{\em The Astrophysical Journal}, 218, 148, 1977

\bibitem[\protect\citeauthoryear{Melioli et al.}{2009}]{melioli}
C. Melioli, F. Brighenti, A. D'Ercole and E. M. De Gouveia Dal Pino.
\newblock{Hydrodynamical simulations of Galactic fountains II: evolution of multiple fountains.}
\newblock{\em MNRAS}, 399, 1089, 2009

\bibitem[\protect\citeauthoryear{Milogradov-Turin \& Smith}{1973}]{Milogradov}
J. Milogradov-Turin and F. G. Smith.
\newblock{A survey of the radio background at 38 MHz.}
\newblock{\em MNRAS}, 161, 269, 1973

\bibitem[\protect\citeauthoryear{Naselsky et al.}{2010}]{cold_spot_env}
P. D. Naselsky, et al.
\newblock{Understanding the WMAP Cold Spot mystery.}
\newblock{\em Astrophys. Bull.}, 65, 101, 2010 

\bibitem[\protect\citeauthoryear{Oppermann et al.}{2011}]{Faraday_depth}
N. Oppermann, et al.
\newblock{An improved map of the Galactic Faraday sky.}
\newblock {\em arXiv:1111.6186.}, 2011

\bibitem[\protect\citeauthoryear{Oppermann et al.}{2011}]{Faraday-map0}
N. Oppermann, H. Junklewitz, G. Robbers and T. A. Ensslin.
\newblock{Probing magnetic helicity with synchrotron radiation and Faraday rotation.}
\newblock{\em Astronomy \& Astrophysics}, 530, a89, 2011.

\bibitem[\protect\citeauthoryear{Quigley \& Haslam}{1965}]{Quigley}
M. J. S. Quigley  and C. G. T. Haslam.
\newblock{Structure of the Radio Continuum Background at High Galactic Latitudes.}
\newblock{\em Nature}, 208, 741, 1965

\bibitem[\protect\citeauthoryear{Rudnick}{2007}]{rudnick}
L. Rudnick, S. Brown, and L. R. Williams.
\newblock{Extragalactic Radio Sources and the WMAP Cold Spot.}
\newblock{\em The Astrophysical Journal}, 671, 40, 2007 

\bibitem[\protect\citeauthoryear{Smith \& Huterer}{2010}]{smith_huterer}
K.M. Smith and D. Huterer.
\newblock{No evidence for the cold spot in the NVSS radio survey.}
\newblock{\em MNRAS}, 403, 2, 2010 

\bibitem[\protect\citeauthoryear{Stil}{2009}]{stil}
J. M. Stil, M. Krause, R. Beck and A. R. Taylor.
\newblock{The Integrated Polarization of Spiral Galaxy Disks.}
\newblock{\em The Astrophysical Journal}, 693, 1392, 2009

\bibitem[\protect\citeauthoryear{Urosevic}{2002}]{Urosevic}
D. Urosevic.
\newblock{Empirical $\Sigma$ - D relations and main galactic radio loops.}
\newblock{\em Serb. Astron. J.}, 165, 27, 2002 

\bibitem[\protect\citeauthoryear{Verkhodanov \& Khabibullina}{2010}]{Mosaic}
O.V. Verkhodanov and M.L. Khabibullina.
\newblock{Dominant Multipoles in WMAP5 Mosaic Data Correlation Maps.}
\newblock{\em Bull. Spec. Astrophys. Obs.}, 65, 390, 2010

\bibitem[\protect\citeauthoryear{Vielva et al.}{2004}]{vielva2004}
P. Vielva, E. Martinez-Gonzalez, R. B. Barreiro, J. L. Sanz and L. Cayon.
\newblock{Detection of non-Gaussianity in the WMAP 1-year data using spherical wavelets.}
\newblock{\em The Astrophysical Journal}, 609, 22, 2004 
\end{thebibliography}
\end{document}